\documentclass[fleqn,usenatbib]{mnras}

\usepackage{newtxtext,newtxmath}

\usepackage[T1]{fontenc}

\DeclareRobustCommand{\VAN}[3]{#2}
\let\VANthebibliography\thebibliography
\def\thebibliography{\DeclareRobustCommand{\VAN}[3]{##3}\VANthebibliography}

\usepackage{graphicx}
\usepackage{xcolor}
\usepackage{rotating}
\usepackage{amsmath,graphics}
\usepackage[utf8]{inputenc}

\newcommand{\Msun}{\rm{M}_{\odot}}

\def\be{\begin{equation}}
\def\ee{\end{equation}}

\def\bea{\begin{eqnarray}}
\def\eea{\end{eqnarray}}

\title[Evidence of a population of dark subhalos]{Evidence of a population of dark subhalos from Gaia and Pan-STARRS observations of the GD-1 stream}

\author[N. Banik et al.]{
Nilanjan Banik,$^{1,2,3}$\thanks{E-mail: banik@tamu.edu}
Jo Bovy,$^{4}$
Gianfranco Bertone,$^{2}$
Denis Erkal,$^{5}$
and T.J.L. de Boer $^{6}$
\\
$^{1}$Mitchell Institute for Fundamental Physics and Astronomy, Department of Physics and Astronomy, Texas A\&M University, College Station, TX 77843, USA\\
$^{2}$GRAPPA Institute, Institute for Theoretical Physics Amsterdam \\
and Delta Institute for Theoretical Physics, University of Amsterdam,Science Park 904, 1098 XH Amsterdam, The Netherlands \\
$^{3}$Lorentz Institute, Leiden University, Niels Bohrweg 2,Leiden, 2333 CA, The Netherlands \\
$^{4}$ Department of Astronomy and Astrophysics, University of Toronto, 50 St. George Street, Toronto, ON, M5S 3H4, Canada \\
$^{5}$ Department of Physics, University of Surrey, UK \\
$^{6}$ Institute for Astronomy, University of Hawai`i, 2680 Woodlawn Drive, Honolulu, HI 96822, USA 
}

\date{Accepted XXX. Received YYY; in original form ZZZ}

\pubyear{2020}

\begin{document}
\label{firstpage}
\pagerange{\pageref{firstpage}--\pageref{lastpage}}
\maketitle

\begin{abstract}
New data from the {\it Gaia} satellite, when combined with accurate photometry from the Pan-STARRS survey, allow us to accurately estimate the properties of the GD-1 stream. Here, we analyze the stellar density variations in the GD-1 stream and show that they cannot be due to known baryonic structures like giant molecular clouds, globular clusters, or the Milky Way's bar or spiral arms. A joint analysis of the GD-1 and Pal 5 streams instead requires a population of dark substructures with masses $\approx 10^7$ to $10^9 \ \Msun$. We infer a total abundance of dark subhalos normalised to standard cold dark matter $n_{\rm sub}/n_{\rm sub, CDM} = 0.4 ^{+0.3}_{-0.2}$ (68\%), which corresponds to a mass fraction contained in the subhalos $f_{\rm{sub}} = 0.14 ^{+0.11}_{-0.07} \%$, compatible with the predictions of hydrodynamical simulation of cold dark matter with baryons.

\end{abstract}

\begin{keywords}
Cosmology: dark matter --- Galaxy: evolution --- Galaxy: halo --- Galaxy: kinematics and dynamics --- Galaxy: structure
\end{keywords}

\section{Introduction}

A central prediction of the standard cold dark matter (CDM) paradigm is that a very large number of dark matter substructures exist inside galactic halos, with masses smaller, possibly by many orders of magnitude, than that of dwarf galaxies \citep{Diemand2008,Springel2008}. 
Detecting these subhalos would confirm a key prediction of standard cosmology and provide crucial hints on the nature of dark matter ~\citep{Bertone:2010zza,Jungman:1995df,Bergstroem2000,Bertone05}. It would in particular rule out alternative models that lead to a suppression of primordial density fluctuations on small scales, such as the so-called warm dark matter models (WDM) \citep{1982ApJ...258..415P} or models where dark matter cannot cluster on small scales, as in the case of ultralight scalars \citep{Hui2016}. Subhalos in this regime are hard to study observationally, because they are dark matter dominated and have very few, if any, stars. Interesting constraints however arise from Ly$\alpha$ forest observations \citep{Narayanan:2000tp,Viel2005,Boyarsky:2008xj,Viel2013,Baur:2015jsy,Garzilli:2015iwa,Irsic:2017ixq}, the study of perturbations in strong gravitational lensing systems \citep{Dalal2002,Vegetti:2008eg,Li:2015xpc,Penarrubia:2017nzw,Asadi:2017ddk,Mao2017,Daylan2017,Minor:2016jou,Despali:2016meh}, and satellite counts around the Milky Way \citep{maccio_fontanot_2010,polisensky_ricotti_2011,Lovell2013,jethwa_etal_2018,kim_2018,nadler_2019}.

Stringent complementary constraints can be obtained from the analysis of the perturbations induced by sub-dwarf dark matter clumps on {\it stellar streams}. Stellar stream originate from the tidal disruption of globular clusters or dwarf galaxies merging into the Milky Way, and exhibit an elongated, almost one-dimensional structure with rather uniform stellar density  \citep{Johnston1998,Sanders2013,Bovy2014}. When a dark subhalo gravitationally perturbs a stream, the long-term effect is that it pushes stars in the stream away from the point of closest approach and thus creates a characteristic gap in the density distribution of stream stars \citep{Yoon2011,Carlberg2012,Carlberg2013,Erkal2015,Erkal2015a,Sanders2016}. Because streams are perturbed by the entire population of dark subhalos, the signatures of different impacts overlap and generically lead to a complicated pattern of density fluctuations \citep{Bovy2016a}. By analyzing the power spectrum of density fluctuations in a stream, one can go beyond the study of individual gaps: dark subhalos of a given mass give rise to density fluctuations on and above a certain scale, with lower-mass, smaller subhalos affecting smaller scales along the stream; the power spectrum therefore encodes the mass function of dark subhalos \citep{Bovy2016a}.

Here, we focus on the GD-1 stream \citep{Grillmair2006}, and make use of data from {\it Gaia} DR2~\citep{GAIAmain1,GAIAmain2,Lindegren18}, combined with accurate photometry from the Pan-STARRS survey data release 1~\citep{Chambers16}, to obtain a sample of stars with both accurate proper motions as well as accurate photometry \citep{Boer2019}. 
A similar combination of data was recently used to characterize the stellar distribution in GD-1 \citep{Price-Whelan18a,Webb2018}, to highlight the existence of stream members that are off the main stream track, and to argue that the observed morphology of off-track stars are probably due to perturbation from dark matter substructures in the Milky Way and can be used to constrain them \citep{Bonaca2018}. 

We perform a full density power spectrum analysis of the normalised density profile as a function of angle along the GD-1 stream obtained in \citet{Boer2019} following the procedure of \citet{Bovy2016a}. We study the effects due to the baryonic substructures; namely, the bar,  spiral arms, giant molecular clouds (GMCs), and the Milky Way's globular clusters (GCs) on the GD-1 stream. We demonstrate that the observed density variations cannot be due to the baryonic structures alone, which strongly hints at the existence of a population of dark substructures. By modeling the combined effect of baryonic and dark substructures, we show that the abundance of dark subhalos required to account for the observed level of density fluctuations is $0.7 ^{+0.9}_{-0.5}$ times a fiducial CDM abundance at 68\% and $< 2.7$ times the fiducial CDM abundance at 95\%, which matches the predictions of the CDM paradigm. We then apply the same analysis to data on the Pal 5 stream whose stellar density data is obtained from \citet{Ibata2016}. The Pal 5 stream, due to its passage through the Galactic disk close to the Galactic center is severely perturbed by the bar \citep{Erkal2017,Pearson2017,Banik2019}, the GMCs \citep{Amorisco2016,Banik2019} and the spiral arms \citep{Banik2019}. As such, it is difficult to detect the influence of dark subhalos on Pal 5, but we demonstrate that Pal 5's observations limit the abundance of dark substructures to be $< 0.9$ times the fiducial CDM abundance at 95\% confidence. Finally, we combine the constraints of both streams to obtain a joint posterior on the abundance of dark substructures within a Galactocentric radius of 20 kpc, which yields $0.4 ^{+0.3}_{-0.2}$ times a fiducial CDM abundance at 68\% and $< 0.9$ times the fiducial CDM abundance at 95\%.

The paper is organised as follows: in Sec.~\ref{sec:GD-1}, we introduce the GD-1 stream, describing the stream data in Sec.~\ref{sec:data} and our modelling of the GD-1 stream in Sec.~\ref{sec:GD-1 model}; in Sec. \ref{sec:perturbers}, we discuss how we model the baryonic and dark matter perturbations to the GD-1 stream; in Sec. \ref{sec:constraints}, we first briefly introduce the Pal 5 stream and how we obtain its data, and then present the results on the amount of dark substructures based on the analysis of the GD-1 and Pal 5 streams; in Sec. \ref{sec:mwdm} we present the constraints on the mass of a thermal dark matter relic;
finally, in Sec. \ref{sec:conclusions} we discuss our results and present our conclusions. The implications of our results for WDM models in particular are elaborated on further in a companion paper \citep{Banik2020L}.

\section{The GD-1 stream}
\label{sec:GD-1}

First detected in the Sloan Digital Sky Survey (SDSS) Data release 4 photometry, the GD-1 stream was found to span 63$^{\circ}$ in the sky \citep{Grillmair2006}. Later in SDSS DR7 the stream was found to span nearly 70$^{\circ}$ in the sky \citep{Willett2009,Koposov2010}. Although no progenitor of this stream has been detected as yet, its mean transverse width, metallicity, and stellar mass indicate that it originated from a globular cluster. Subsequent follow-up with the Canada-France-Hawaii-Telescope (CFHT) revealed several deep gaps and wiggles in the stream \citep{de_boer_cfht_gd1}. A recent follow up study using astrometric data from \textit{Gaia} DR 2 and photometric data from Pan-STARRS has revealed $20^{\circ}$ more of the stream \citep{Price-Whelan18a,Webb2018} making the GD-1 stream to span nearly 90$^{\circ}$ in the sky. In addition, high-contrast gaps in the stellar distribution along the stream were also found. Thanks to its retrograde orbit as well as its distant passage from the Galactic center with a perigalacticon of $\sim 14$ kpc, the GD-1 stream is expected to be only mildly affected by the baryonic substructures in the disk, making it the ideal stellar stream for probing dark substructures \citep{Amorisco2016}.

\subsection{GD-1 stream data}
\label{sec:data}

To study GD-1, we  make use of the stream properties as presented by \citet{Boer2019}. In that work, data from {\it Gaia} DR2~\citep{GAIAmain1,GAIAmain2,Lindegren18} was combined with accurate photometry from the Pan-STARRS survey, data release 1~\citep{Chambers16} to obtain a sample of stars with both accurate proper motions as well as accurate photometry. 

\begin{figure}
\includegraphics[width=0.5\textwidth]{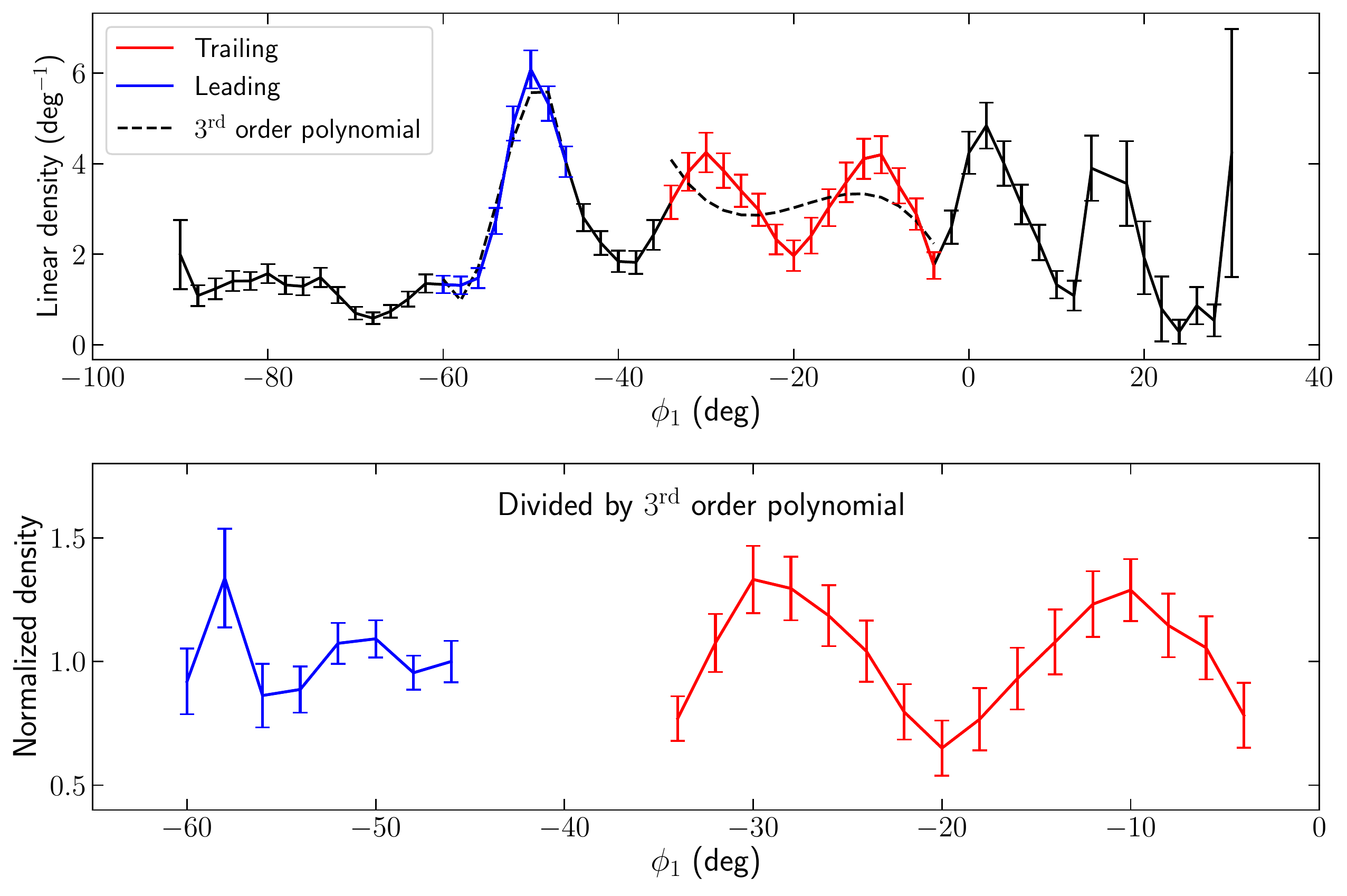}
\caption{GD-1 density data. The top panel shows the linear density of the GD-1 stream as a function of the stream-aligned sky coordinate $\phi_1$, as determined by fitting a Gaussian plus 1$^{\rm{st}}$ order polynomial background to the {\it Gaia} DR2 data. We base our fiducial model on \citet{Webb2018} in which the progenitor disrupted $\sim$ 500 Myr ago and resulted in the gap at $\phi_{1} = -40^{\circ}$. We exclude the underdense region of the stream within $6^{\circ}$ of $\phi_{1} = -40^{\circ}$ that is in the immediate vicinity of the disrupted progenitor and was likely caused by the total disruption of the progenitor. For our analysis, we consider the stream between $-60^{\circ} < \phi_{1} < -4^{\circ}$. Colored in blue and red are the leading and trailing arms that we consider in our analysis. The black dashed curve is the $3^{\rm{rd}}$ order polynomial fit to the density that is used to normalize the stream density which is shown in the bottom panel.}
\label{fig:GD1_dens}
\end{figure}

A matched filter technique (see, e.g., \citet{Rockosi02}) was employed in concordance with newly determined distances to the different parts of the stream to obtain the spatial distribution of GD-1 in the stream-aligned sky coordinate scheme of  \citet{Koposov2010}, with coordinate $\phi_1$ roughly along the path of stream and coordinate $\phi_2$ perpendicular to the stream. This resulted in a detailed normalized density profile as a function of angle along the stream, as well as constraints on the nominal stream track. The linear density profile (see Figure \ref{fig:GD1_dens}) shows that within {\it Gaia} DR2 data, the stream is mostly contained to within $-60^{\circ} < \phi_{1}< -4^{\circ}$, as a result of the {\it Gaia} limiting magnitude. While there are clearly stream stars found beyond those limits, the density is sufficiently low (and the contamination from field stars sufficiently high) that we limit our study of GD-1 to this stream section. Furthermore, the main features of interest (gaps and density variations) are only seen within these proposed limits, making it the obvious region of interest. We note that the linear density profile does not include the spur and blob features presented in \citet{Price-Whelan18a}, which will otherwise convolve the main stream track density with the density of off-stream stars originating from a different stream angle location. The error bars on the density are computed separately for each angle bin based on the uncertainty on density and width of the convolved stream data (using a kernel of 1$\times$1 bin).

Recently, it was pointed out in \citet{Ibata2020} that the incompleteness in Gaia's scanning pattern in DR2 can result in periodic small scale ($\phi_{1} \sim 0^{\circ}.2$) density gaps along GD-1 in the region $-60^{\circ} < \phi_{1} < -40^{\circ}$ that are not intrinsic to the stream and may contribute to the overall density power of the GD-1 stream. While these small scale incompleteness due to \textit{Gaia}'s scanning pattern has been known for sometime \citep{Arenou2018}, see also \citep{Boer2019} for a detailed discussion on this, it is important to note that they will result in density power at very small scales that we not consider in the power-spectrum analysis below (because DM substructure gives observable density variations at larger scales). We will see that on these very small scales, the power in the density variations in the data is dominated by noise and we therefore exclude them from our analysis (see figures \ref{fig:Pk_baryon_struct} and \ref{fig:Pk_CDM_baryonicstruct} below).

\subsection{Modelling the mock GD-1 stream}
\label{sec:GD-1 model}
We model the GD-1 stream using the frequency-angle ($\Omega,\theta$) framework following \citet{Bovy2014} in the static Milky Way potential \texttt{MWPotential2014} \citep{Bovy2015} \footnote{The generation of mock streams and their evolution are done using the Python package for galactic dynamics \texttt{galpy} \citep{Bovy2015}, available at \url{https://github.com/jobovy/galpy}, and related tools that come with it.}.
This method relies on the phase space coordinates of the progenitor star cluster, the mean velocity dispersion $\sigma_{v}$ of its member stars, and the time in the past when it started to disrupt $t_{d}$. In forming the stream we have assumed a constant stripping rate of the stars from its progenitor cluster throughout its orbit. Realistically this is not accurate since more stars are released during the pericentric phase of the cluster's orbit. N-body simulations performed in \citet{Kuepper2009,Kuepper2011} show that such episodic stripping coupled to the epicyclic motion of the progenitor can lead to periodic regions of over and under density in the stellar density along the tidal tails resembling the observed gaps.  However this phenomenon is mostly confined to regions closest to the progenitor where the stream is youngest and the stars have not mixed enough \citep{Ngan2013,Bovy2016a}. In regions away from the progenitor, these periodic structures are washed out as stars mix over the dynamical evolution of the stream. Since we only focus on the older parts of the stream that are away from the progenitor where the effects of dark matter subhalo impacts are most pronounced, our results are not affected by this assumption. 

The location of the GD-1 progenitor is as yet unknown. However, recent N-body simulations of the GD-1 stream \citep{Webb2018} suggests that the progenitor is likely between $-45^{\circ} < \phi_{1} < -30^{\circ}$ and that it either completely disrupted $\sim 2.5$ Gyr ago leaving no observable signatures or it disrupted only $\sim 500$ Myr ago and resulted in the underdensity at $\phi_{1} \sim -40^{\circ}$. Since there is a clear underdensity in the observed GD-1 stream at $\phi_{1} \sim - 40^{\circ}$, we consider the latter scenario as our fiducial GD-1 model. In \citet{Bonaca2018}, an alternate GD-1 model was suggested in which the progenitor disrupted $\sim 500$ Myr ago and resulted in the gap at $\phi_{1} = -20^{\circ}$. In appendix \ref{sec:prog-20}, we investigate this model and explore how the density power spectra of the leading and trailing arms are affected in this scenario. 

The best fit phase space coordinate of a point along the stream's orbit near the leading end of the GD-1 stream was obtained in \citet{Webb2018}. We compute the phase space coordinate of the progenitor by integrating this point back in time in the \texttt{MWPotential2014} until it reached the observed sky coordinate of $\phi_{1} = -40^{\circ}$. The resulting phase space location of the point is : 
\begin{align*}
\mathrm{RA} &= \phantom{-}148^{\circ}.91 \\
\mathrm{Dec} &= \phantom{-}36^{\circ}.15 \\
D &= \phantom{-}7.56 {\rm ~ kpc} \\
\mu_{\alpha}\cos\delta &= -5.33 {\rm ~mas~yr^{-1}} \\
\mu_{\delta} &= -12.18 {\rm ~mas~ yr^{-1}} \\
V_{\rm{los}} &= \phantom{-}7.16 ~{\rm ~km~ s^{-1}}
\end{align*}
 which we use as the fiducial GD-1 stream progenitor's current phase space coordinates. This method of obtaining the progenitor's location is based on the assumption that the observed stream follows a single orbit which is not generally true (e.g., \citep[e.g.,][]{Eyre2011,Sanders2013a,Bovy2014}). However, for the GD-1 stream the difference between the stream and the orbit is small (e.g., \citep[e.g.,][]{Sanders2013a,Bovy2016}) and our assumption has no effect on the results. The age of the GD-1 stream is unknown. In the N-body simulations of the  GD-1 stream in \citet{Webb2018}, a dynamical age of 3.4 Gyr was found to produce a stream that matches the observed stream's location, overall width and length. However, there are locations along the stream such as in the range $-32^{\circ} < \phi_{1} < -12^{\circ}$ where the Gaussian width is as low as $\sim 6'$. Since the stream width is proportional to $\sigma_{v}$ and the stream length is proportional to $\sigma_{v}\times t_{d}$, presence of regions of such thin stream width could imply that the original stream is much older and that the broader regions of the stream are a result of heating due to impacts with dark matter substructures and/or baryonic structures over the course of its evolution. We do not pursue this further in this paper, instead we follow a conservative approach and marginalize over stream ages of [3,4,5,6,7] Gyr uniformly. For a 3 Gyr old stream we set $\sigma_{v} = 0.32\,{\rm ~km~ s^{-1}}$ which produces a stream that matches the observed stream length and has a mean Gaussian width of $\sim 12'$ over the stream region $-60^{\circ} < \phi_{1} < -10^{\circ}$, which is consistent with the observed mean stream width. This stream width is also consistent with past works such as \citep[e.g.,][]{Koposov2010,Carlberg2013a,de_boer_cfht_gd1}. In order to keep the stream length fixed for older stream models, we adjust $\sigma_{v}$ as $(3\,{\rm Gyr}/t_d)\times 0.32\,{\rm ~km~ s^{-1}}$ which yields a mean Gaussian width of $5.7'$ for a 7 Gyr old stream, that is consistent with the width of the thinnest regions of the stream. We remove $6^{\circ}$ to the left and right of $\phi_{1} = -40^{\circ}$ in order to remove any density variations caused by the disrupting progenitor. The bottom panel in Figure \ref{fig:GD1_dens} shows the normalized linear density of the leading and trailing arm of the GD-1 stream. In Appendix \ref{sec:cut_width}, we investigate how the size of the cut around the progenitor affects the stream density power spectrum.

\subsection{Modeling the effects of the perturbers }
\label{sec:perturbers}

The observed density and track variations along the GD-1 stream indicate that it encountered perturbers over its dynamical age. In this section, we investigate how different perturbers affect the GD-1 stream. Following the formalism laid out for the Pal 5 stream by \citet{Banik2019}, we consider perturbations from the known baryonic substructures namely Galactic bar, the spiral arms, the Galactic population of giant molecular clouds (GMCs), and globular clusters (GCs), and the unknown dark matter substructures, which we aim to constrain in this work. In all cases, we quantify the density variations at different angular scales of the stream by computing the stream density power spectrum following the same procedure as in \citet{Bovy2016a}. 

\subsubsection{Modelling the Baryonic Structures}

Stellar streams, in particular the Pal 5 stream, have been shown to be severely perturbed by the bar \citep{Erkal2017,Pearson2017,Banik2019}, the spiral arms \citep{Banik2019} and the GMCs \citep{Amorisco2016,Banik2019}. The GD-1 stream has a perigalacticon of 13.5 kpc and is in a retrograde orbit making it much less susceptible to perturbations from the bar, spiral arms, and GMCs. In this subsection, we first carry out a detailed analysis of the effects of the bar and the the spiral arms, and then we explore the effects of the GMCs and the GCs on the GD-1 stream. Finally, we combine their effects to statistically estimate how much they can perturb the GD-1 stream density.

\paragraph{Bar.} We use the fiducial bar model from \citet{Banik2019} which has a triaxial, exponential density profile following \citet{Wang2012}, a mass of $10^{10}~ \rm{M}_{\odot}$, rotating with a pattern speed of 39 km s$^{-1}$kpc$^{-1}$ \citep{Portail2016,Bovy2019,sanders_bar} and 5 Gyr old. Recently, \citet{Bovy2019} found that the bar is likely older, $\sim 8$ Gyr. However, based on figure 9 in \citet{Banik2019} where it was shown that varying the age of the bar over [2,3,4,5] Gyr did not change the bar's effect on the density power spectrum of the Pal 5 stream much, it is a good assumption that the effect of an 8 Gyr old bar will not be too different from a 5 Gyr old one. We follow the same procedure for incorporating it in the \texttt{MWPotential2014} by replacing the bulge of mass $5\times 10^{9} \ \Msun $ by the bar and removing the mass equal to the extra mass of the bar from the disk. This ensures the total baryonic mass of our Milky Way model stays constant. We set the present day angle of the bar’s major axis with respect to the Sun–Galactic-centerline to 27$^{\circ}$ \citep{Wegg2013}. 

\paragraph{Spiral arms.} The spiral arms are modeled using the analytic expression of its potential from \citet{Cox2002}. A four arm spiral whose density amplitude is such that it corresponds to 1\% of the total radial force at the location of the Sun was shown to cause the most perturbation to the Pal 5 stream in \citet{Banik2019}. In the light of this result, we use the same four arm spiral in this study. In particular, we set the pattern speed of the spiral arms to 19.5 km s$^{-1}$kpc$^{-1}$, the radial scale length to 3 kpc which is similar to the disk scale length of the \texttt{MWPotential2014}, and the vertical scale height is set to 0.3 kpc. Following \citet{Siebert2012,Faure2014,Monari2016a}, we set the pitch angle of the spiral structure to $9.9^{\circ}$ and the reference angle to $26^{\circ}$. We add the spiral potential to the barred Milky Way potential and construct an effective Milky Way potential with a bar and spiral arms. Beside the steady state spiral arms, we also tested the effect of transient spiral structure by modulating the amplitude of the steady-state spiral with a Gaussian (such that there are $\sim 6$ spiral episodes per Gyr) and find a similar amount of power as in the steady-state spiral arms case.

\begin{figure}
\includegraphics[width=0.5\textwidth]{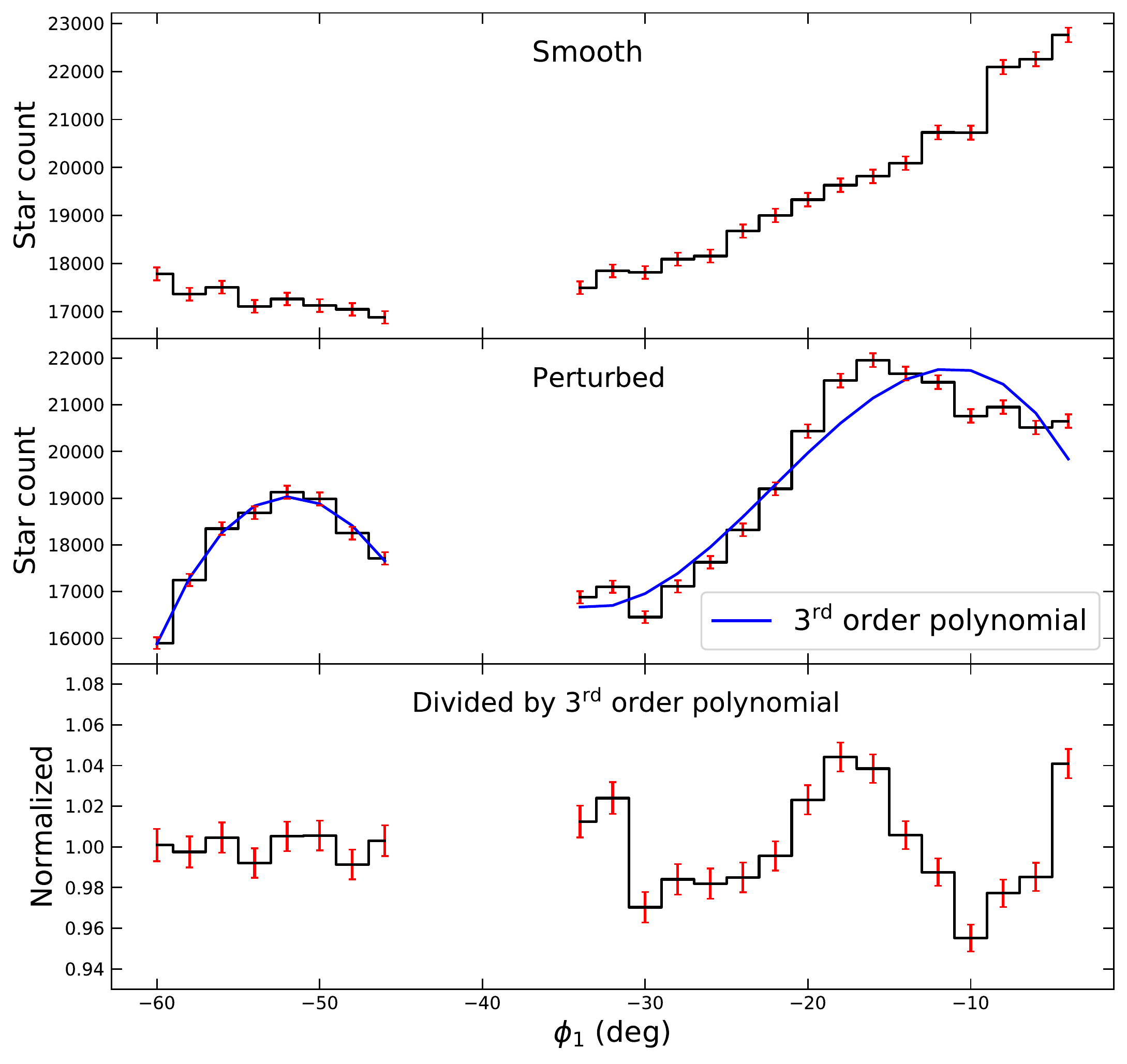}
\caption{Star count distribution along a mock GD-1 stream with an age of 5 Gyr. The top panel shows the case of the smooth stream that was evolved in the axisymmetric Milky Way potential, the middle panel shows the case where the stream was evolved in the Milky Way potential with an added bar and spiral arms. The blue 3$^{rd}$ order polynomial is obtained by fitting the star counts and is divided out to give the normalized stream density which is shown in the bottom panel. The error bars in each panel is the binning shot noise. The star count rises towards $\phi_{1} = -10^{\circ}$ due to projection effects while transforming from $\theta_{\parallel}$ to $\phi_{1}$ coordinates (Jacobian of transformation). Overall, the effect of the bar and spiral arms on the star counts along the GD-1 stream is small.}
\label{fig:nden_bar_spiral}
\end{figure}

We evolve a mock GD-1 stream in the above potential using a combination of the the frequency-angle framework and orbit integration. In practice, we first generate the mock stream in the axisymmetric Milky Way potential and sample the phase space coordinates today of $10^6$ points and their corresponding time when they were stripped from the progenitor. We then integrate each point back in the axisymmetric potential until their respective time of stripping and finally, we integrate them forward in the bar + spiral Milky Way potential until today. The final phase space coordinates of the points are transformed to the custom sky coordinates $(\phi_{1},\phi_{2})$ of the GD-1 stream. We select the points that are in the range $-60^{\circ} < \phi_{1} < -4^{\circ}$, excluding the ones between $-46^{\circ} < \phi_{1} < -34^{\circ}$ to remove effects of the disrupting progenitor. We bin these points in $2^{\circ}$-wide $\phi_{1}$ bins and then fit a third order polynomial through them. We divide the bin counts by this polynomial to suppress large scale variations which could stem from factors such as non-constant stripping rate which we have assumed to be constant in our model. We stress that in the polynomial normalization we divide the stream density by a smoothing polynomial and it is not meant to fit the data perfectly. As was done in \citet{Bovy2016a}, we have checked the effect of using a $2^{\mathrm{nd}}$ and $4^{\mathrm{th}}$ order smoothing polynomial and found the final results to be consistent with each other. In Figure \ref{fig:nden_bar_spiral}, we show the binned distribution of the sampled points for a 5 Gyr old GD-1 stream in the axisymmetric Milky Way potential (top panel), in the bar+spiral Milky Way potential (middle panel), and the normalized density (bottom panel). The error bars are due to binning shot noise. As evident from the normalized density, the bar and spiral arms do not affect the stream density of the GD-1 stream significantly. In this method, the deviations of the progenitor's orbit in the bar+spiral Milky Way potential compared to the axisymmetric Milky Way potential is neglected. However, following the same steps as in \citet{Banik2019}, we have checked that including the perturbations to the progenitor's orbit has no observable effect on the stream density.  

\paragraph{GMCs, GCs and classical Milky Way satellites}
\label{sec:GMC_GC}

We included the effects of the GMCs and GCs on the GD-1 stream by again following the same steps as in \citet{Banik2019}. We briefly describe the steps here. We use the position and velocity of the GMCs from the GMC catalog from \citet{Miville-Deschenes2016} which we then correct for empty patches in the outer disk on the other side of the Galactic center. The GMCs are modeled as Plummer spheres of scale radius equal to one-third of their observed radius since most of the mass is contained within this radius. We consider only the GMCs that are at least $10^{5}$ M$_{\odot}$, since GMCs less massive than that have no observable effects on the stellar streams. We set them on circular orbits based on their Galactocentric radial distance in the \texttt{MWPotential2014}. We then integrate their orbits back in the same potential until the dynamical age of the stream and then evolve both the mock GD-1 stream and the GMCs forward until today. During this evolution, the encounters between the GMCs and the stream are approximated using the impulse approximation and the final stream density is computed using the same line-of-parallel-angles method that we use for computing the effect of dark subhalos below, as described in \citet{Bovy2016a}. The typical lifetime of a GMC is $10 - 50$ Myr \citep{Jeffreson2018} and so the GMC population has evolved substantially over the dynamical age of the GD-1 stream. To incorporate the effects of the evolving population of the GMCs, we add random rotations to the Galactocentric $\phi$ coordinates of the present day population of GMCs before rewinding them back in time. We then run many realizations of the interactions of the GMCs with the stream and study the resulting stream densities statistically. 

We treat the GCs as Plummer spheres as well and take their sky and kinematic coordinates, mass, and size information of 150 GCs from the updated catalog from \citet{Vasiliev2018}. However, instead of assigning them their mean proper motions and line of sight velocities from the catalog, we sample Gaussian random noise from within their kinematic uncertainties and add them to their mean proper motions and line of sight velocities. This allows us to explore the range of possible orbits of the GCs and consequently their effects on the stream density. We then integrate the GCs back in \texttt{MWPotential2014} until the age of the stream and then integrate them forward with the stream, computing their effects on the latter. Similar to the GMCs, we run many realizations of the GCs interacting with the stream and study the stream density statistically. In practice, we load both the GMC and GC encounters and compute their effects together in our simulations. 

The line-of-parallel-angles method used to compute the effects of the GMCs and GCs is based on the frequency-angle framework as developed in \citet{Bovy2016a}. This framework only supports axisymmetric potentials and so we can not include the bar + spiral Milky Way potential in the simulations of the GMCs and GCs. To compute the cumulative effects of the bar, spiral arms, GMCs, and GCs we add the density perturbation due to the bar + spiral Milky Way potential to the perturbed density due to the GMC and GC impacts. The density perturbation due to the bar + spiral Milky Way potential is the result of subtracting the smooth density from the perturbed density. In Appendix \ref{sec:add densities} we show that this method works, as expected if at least one of the two sets of perturbations is small (and, thus, close to linear). We then normalize the total perturbed density by fitting a $3^{\rm{rd}}$ order polynomial to it and then dividing the density by it. Following the same steps as in \citet{Bovy2016a}, we use the normalized density to compute the power spectrum. 

\begin{figure*}
\includegraphics[width=0.8\textwidth]{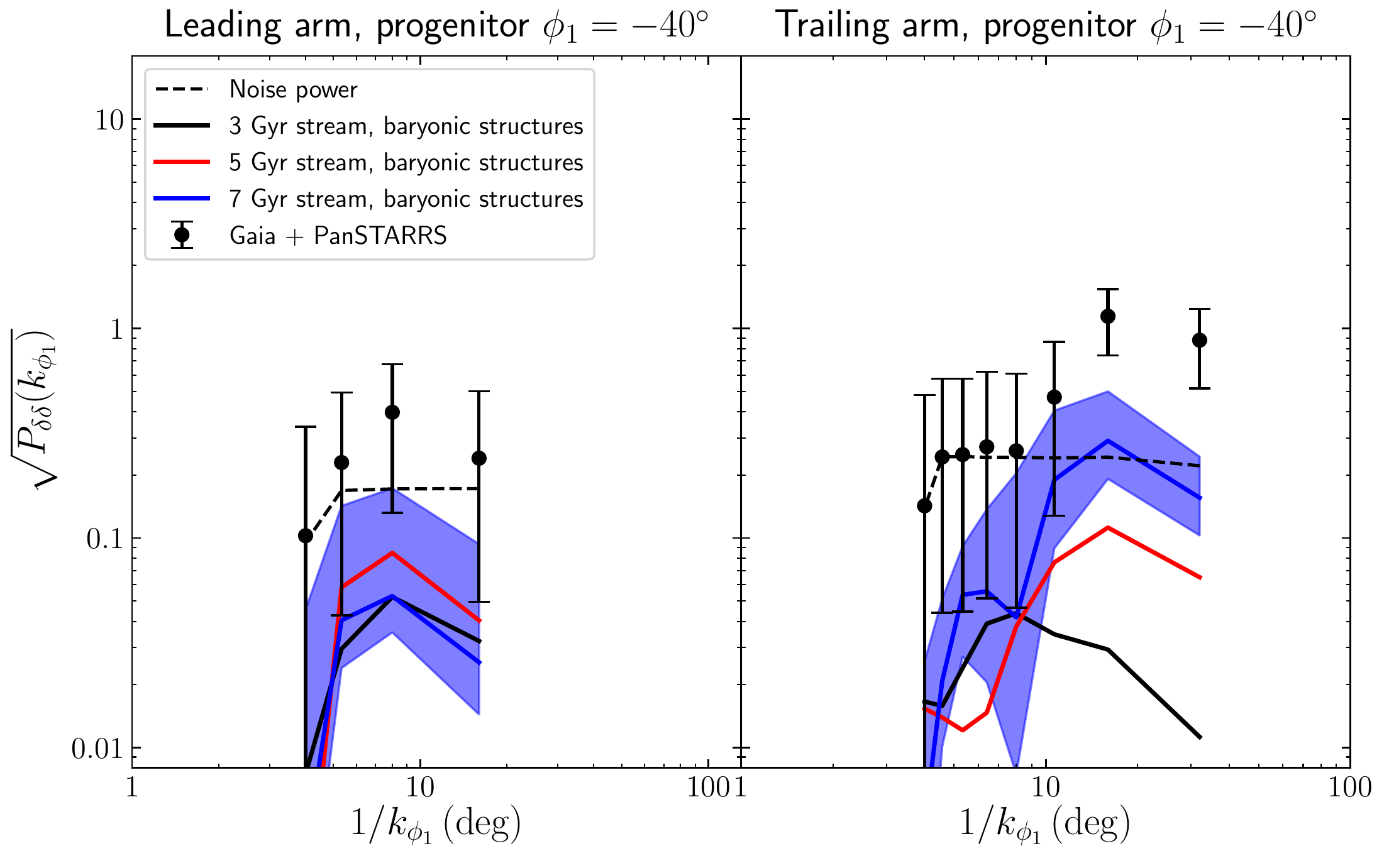}
\caption{Density power spectrum of the leading and trailing arm of the GD-1 stream compared with that of mock GD-1 streams of different ages perturbed only by baryonic substructures. The black dots represent the median density power of the observed stream and the errorbars denote the $2\sigma$ scatter due to the noise in the observed density. The black dashed line represents the median power of the noise in the density. The solid lines represent the median density power of 1,000 mock simulation runs with different age of the stream with the blue line representing the case for a 7 Gyr old stream, the red for a 5 Gyr old stream, and black a 3 Gyr old stream. The blue shaded region represents the $2\sigma$ scatter of the density power for the 7 Gyr old stream. Baryonic substructure alone cannot account for the density fluctuations observed in GD-1 on large scales.}
\label{fig:Pk_baryon_struct}
\end{figure*}

In Figure \ref{fig:Pk_baryon_struct}, we show the density power spectrum of the GD-1 stream due to the perturbations imparted by all of the baryonic structures for three different ages of the stream. The black points show the median power of the observed GD-1 stream density. The error bars denote the $2\sigma$ scatter around the median power due to the noise in the density data that was assumed to be Gaussian. The dashed line shows the median power of the density noise. The blue, red, and black solid lines show the median density power of 1,000 mock GD-1 stream realizations that are 7, 5 and 3 Gyr old respectively. The blue shaded region shows the $2\sigma$ scatter of density power for the 7 Gyr mock stream case. The $2\sigma$ scatter for the other cases are of similar width. There is a consistent rise in power on large scales of the trailing arm as the age of the stream is increased. This is expected since an older stream has more time to interact with the baryonic structures and consequently gets more perturbed by them. For the leading arm, since most of the density information is removed due to the cut around the progenitor, the small amount of power caused by the baryonic structures does not strongly depend on the age of the stream and the dispersion of the density power are all within each other's scatter. As evident from this figure, baryonic structures alone can not account for the observed density power of the stream.    

Finally, we analyzed the cumulative effect of all the classical Milky Way satellites by running 1000 simulations of their gravitational encounters. For each simulation we took the current 6D phase space information of the satellites from \citet{Fritz2018} and sampled their proper motions, line of sight velocity and distance from their uncertainties. These were then integrated back for 3 Gyr in \texttt{MWPotential2014} and then integrated forward until today along with the GD-1 stream and their impacts on the stream’s density were computed in the same way as the GMCs and GCs. We assigned the satellites' mass using the halo abundance matching relation from \citet{2017ARA&A..55..343B} and their scale lengths using the same fitting relation that was used for the GMCs. Doing so we found that the satellites imparted density power at the level of $\sim 10^{-2}$ and therefore insignificant compared to the perturbations from other baryonic structures, and have therefore not included them in the subsequent analyses in this paper.

\subsubsection{Modeling the dark matter subhalos}
\label{sec:subhalos}

We model cold and warm 
dark matter subhalos in our simulations following our previous works \citep{Bovy2016a,Erkal2016,Banik2018}, which we briefly describe here. We are interested in subhalos that are in the sub-dwarf-galaxy mass range $[10^{5} - 10^{9}] \ \Msun$. Subhalos less massive than $10^{5} \ \Msun$ have no currently-observable effect on the stream density. For our fiducial CDM model, we use the mass function $dN/dM \propto M^{-1.9}$ and consider their radial distribution inside the Milky Way to follow an Einasto profile following \citet{Springel2008}. \citet{Erkal2016} combined these results to obtain a normalized subhalo profile for a Milky Way sized host galaxy to be 

\begin{equation}
\left(\frac{dn}{dM}\right)_{\rm{CDM}} = c_{0}\left(\frac{M}{m_{0}}\right)^{\alpha}\exp\left\{ - \frac{2}{\alpha_{r}}\left [\left(\frac{r}{r_{-2}}\right)^{\alpha_{r}} - 1 \right]\right\}
\label{eq:dndMc}
\end{equation}
where the amplitude $c_{0}=2.02 \times 10^{-13}~\Msun^{-1}$~kpc$^{-3}$, slope $\alpha = -1.9$, $m_{0}= 2.52\times 10^{7}~\Msun$, $\alpha_{r} = 0.678$ and $r_{-2} = 162.4$~kpc. We use this profile and amplitude as our fiducial CDM prediction.

 We use the WDM mass function from \citet{Lovell2013} that was obtained by fitting WDM subhalos within a Milky Way like host galaxy in a high resolution N-body simulation based on the Aquarius project \citep{Springel2008}

\begin{equation}
\left(\frac{dn}{dM}\right)_{\rm{WDM}} =  \left(1+ \gamma\frac{M_{\rm{hm}}}{M}\right)^{-\beta} \left(\frac{dn}{dM}\right)_{\rm{CDM}},
\label{eq:wdmcdm}
\end{equation}
where $\gamma = 2.7$ and $\beta = 0.99$. The half-mode mass denoted by $M_{\rm{hm}}$ is the threshold mass below which the mass function is strongly suppressed. It is equal to the mean mass contained within a radius of half-mode wavelength that is defined as $\lambda_{\rm{hm}} = 2\pi \alpha_{\rm{cutoff}}(2^{\nu/5} - 1 )^{-1/2\nu}$ with $\nu = 1.12$ \citep{Viel2005} and 

\begin{multline}
\alpha_{\rm{cutoff}} = 0.047\left(\frac{m_{\rm{WDM}}}{\rm{keV}}\right)^{-1.11}\left(\frac{\Omega_{\rm{WDM}}}{0.2589}\right)^{0.11} \\
\times \left(\frac{h}{0.6774}\right)^{1.22}h^{-1}\rm{Mpc},
\end{multline}
where $m_{\rm{WDM}}$ is the WDM particle mass, $\Omega_{\rm{WDM}}$ is DM density parameter, and $h$ is the dimensionless Hubble constant $H_0 = 100\,h\,\mathrm{km\,s}^{-1}\,\mathrm{Mpc}^{-1}$.

\begin{figure*}
\includegraphics[width=0.8\textwidth]{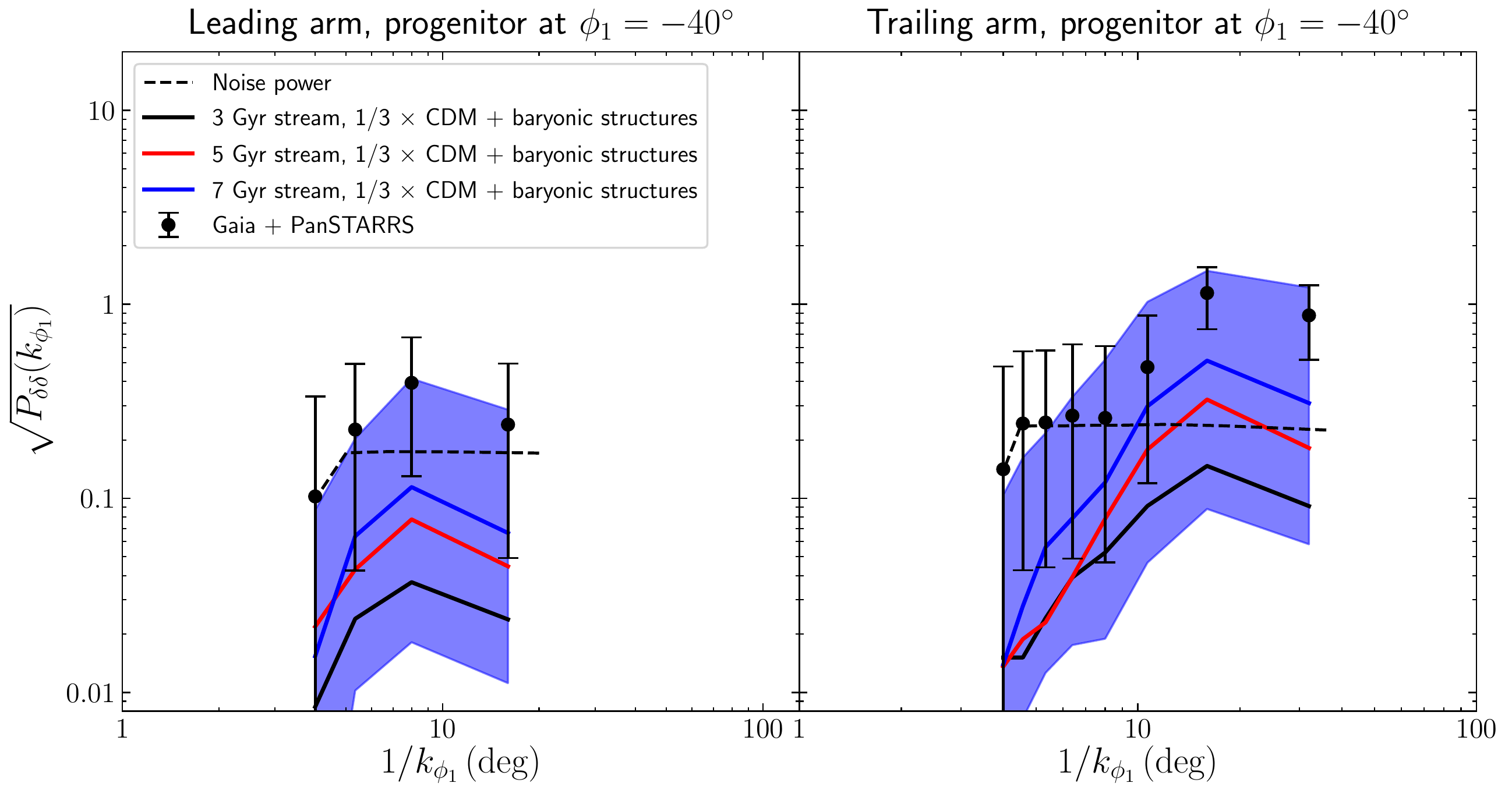}
\caption{Same as Figure \ref{fig:Pk_baryon_struct} but including the effects of impacts by one-third the abundance of subhalos in the fiducial CDM case to take into account subhalo depletion by the baryonic structures. A CDM-like population of dark subhalos in addition to baryonic substructures naturally explains the observed density fluctuations in the GD-1 stream.}
\label{fig:Pk_CDM_baryonicstruct}
\end{figure*}

It is important to note that the above CDM model which we consider as the ``fiducial" model for the rest of the paper is based on dark matter only (DMO) simulations rather than the true CDM prediction which takes into account subhalo disruption by the baryonic structures in the Galaxy. In the latter, around $\sim 10 - 50 \%$ of the subhalos in the mass range $10^{6.5} - 10^{8.5} \Msun$ within the radius of the GD-1 stream is expected to have disrupted \citep{DOnghia10a,Sawala2016,GarrisonKimmel17a,Webb20b}. This percentage is expected to be greater for WDM subhalos due to their lower concentration during their time of accretion. We discuss our results in the light of this expected substructure depletion further below. But we use the DMO prediction as the fiducial, because it is more robustly determined from simulations, since much uncertainty still remains about the exact level of baryonic depletion that is expected. We have also ignored the time evolution of the number density of the subhalo population, since perturbations due to very old impacts are smoothed out due to the velocity dispersion of the stream stars leaving imprints of only the recent impacts visible today.

Having set the mass functions of our dark matter models, we follow the same steps as in Sec 2.3 of \citet{Bovy2016a} to implement the subhalo impacts in the stream simulations. However, in order to prepare ourselves for a later discussion on constraining the sensitivity of different mass subhalos from the observed stream density power, we mention the important points here. The expected number of impacts that a leading or trailing arm will encounter over its lifetime is given by 

\begin{equation}
    N_{\rm{enc}} = \sqrt{\frac{\pi}{2}}r_{\rm{avg}}\sigma_{h}t_{d}^{2}\Delta\Omega^{m}b_{\rm{max}}n_{h}
\end{equation}

where $r_{\rm{avg}}$ is the mean spherical radius of the stream which for our GD-1 model is $\sim$ 20 kpc, $\sigma_{h}$ is the velocity dispersion of the subhalos which we set to 120 km/s, $t_{d}$ is the time since the progenitor star cluster of the stream commenced disrupting, $\Delta\Omega^{m}$ is the mean-parallel-frequency parameter of the smooth stream \citep{Bovy2014}, $b_{\rm{max}}$ is the maximum impact parameter set equal to 5 times the scale radius $r_{s}$ of the subhalo, and $n_{h}$ is the number density of subhalos in the mass range being considered. We describe the subhalos as Plummer spheres with scale radius $r_{s} = 1.62\ \rm{kpc} \ (M_{\rm{sub}}/10^{8}\Msun)^{0.5}$ \citep{Erkal2016} instead of Hernquist spheres. This is simply because Hernquist is a poor description of the GMC profiles and when we simulate the subhalo, GMC, and GC impacts on the stream together, it is more convenient in our simulations to collate impacts from a single type of perturber. This, however, does not affect our results, because it was shown in \citet{Bovy2016a} that at large scales where the signal dominates noise, the density power spectrum of a mock stream impacted by subhalos with a Hernquist profile were indistinguishable from that due to subhalos with a Plummer profile. For each subhalo mass decade, the scale radius $r_{s}$ is computed at its midpoint (e.g., for subhalos in the range $10^{5} - 10^{6} \ \Msun$, $r_s$ is computed at $10^{5.5} \ \Msun$) which is then used to compute the expected number of impacts in each mass decade of the subhalos. The number of impacts the stream encounters is a Poisson draw from the total expected number of impacts. The impact parameter $b$ is sampled from a uniform distribution between $\pm b_{\rm{max}}$ \citep{Erkal2016}. Low mass subhalos need to pass closer to the stream to leave any observable effects, because of this the subhalo mass and its scale radius is drawn from the joint distribution marginalized over the impact parameter $b$: $\int db \ p(M,r_{s},b)$ which translates to the effective distribution $p(M,r_{s}) \propto M^{0.5}dN/dM$. From the sampled subhalo masses their scale radii are computed using the Plummer relation mentioned above. The time of impacts and the angular offsets of the regions of closest approach are computed exactly like in \citet{Bovy2016a}. Having computed all the subhalo impact parameters, they are combined with those of the GMCs and GCs and set to impact the stream using the \texttt{galpy} extension \texttt{streampepperdf}\footnote{Available at \url{https://github.com/jobovy/streamgap-pepper}~.} code. 

Using this approach we ran many simulations of the leading and trailing arm of the GD-1 stream in the fiducial CDM case for the GD-1 stream of age 3, 5 and 7 Gyr. Figure \ref{fig:Pk_CDM_baryonicstruct} shows the resulting power spectra. Like Figure \ref{fig:Pk_baryon_struct}, the solid lines show the median density power of 1,000 realizations in each case. The blue shaded region is the $2\sigma$ scatter of density power for the 7 Gyr case. Running the subhalo impact simulations of each stream arm independent of each other ignores the large scale effects due to the most massive subhalos that affect both arms simultaneously. However, this does not affect our analysis, because we normalize the stream density by a third order polynomial that removes effects from the large scale density variations. When including CDM subhalos in the simulations, especially for older stream models, many realizations encountered so many subhalo impacts that they partially or fully disrupted. While computing the stream density power spectrum we discard those cases by requiring the mock stream length to be at least equal to the observed stream length between $-60^{\circ} < \phi_{1} < -4^{\circ}$, where the mock stream length is defined up to the point where the stream density drops below the $20\%$ of the mean density within $\Delta\phi_{1} = 4^{\circ}$ around the progenitor. For the 5 Gyr old stream $\sim 75\%$ simulations were discarded, while for the 7 Gyr old stream $\sim 99\%$ of the simulations were discarded. The same trend of more density power for older streams is also seen in this case. 

Comparing Figure~\ref{fig:Pk_CDM_baryonicstruct} with Figure \ref{fig:Pk_baryon_struct}, adding the CDM subhalo impacts results in the appropriate density power to account for the observations. That is, a population of low-mass dark subhalos with a number similar to that predicted by CDM can explain the density fluctuations in the GD-1 stream.

\begin{figure}
\includegraphics[width=0.4\textwidth]{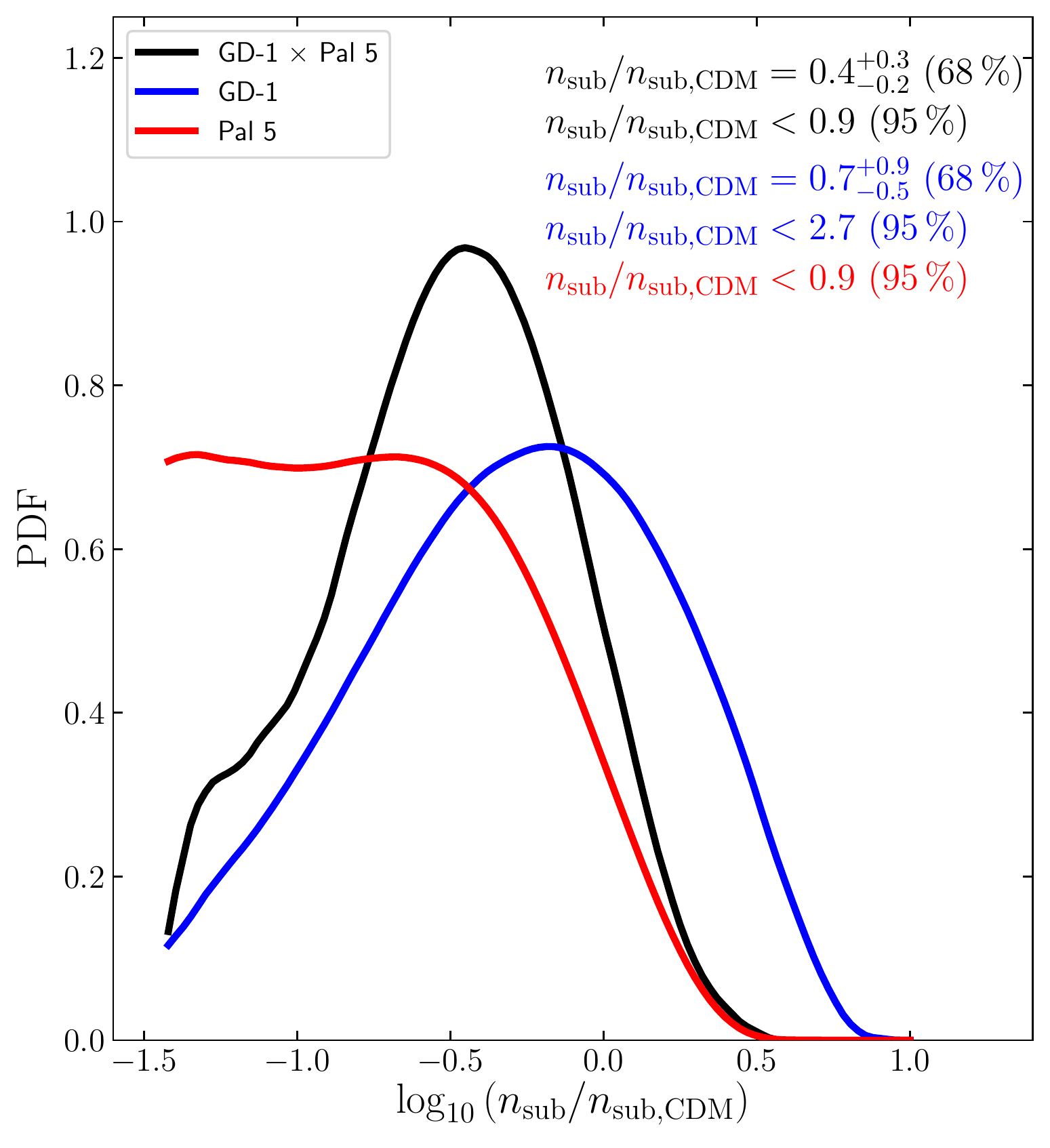}
\caption{Posterior PDF of the abundance of subhalos in the mass range $10^{5} - 10^{9} \ \Msun$ relative to the fiducial CDM case obtained by fitting the GD-1 stream (blue), the Pal 5 stream (red), and both streams simultaneously (black).}
\label{fig:ABC_rate}
\end{figure}

\subsection{The impact of epcicylic overdensities on the observed stream structure}

The smooth stream model that we described in Sec.~\ref{sec:GD-1 model} and that is the basis for our modeling of perturbations to the stream does not include the effect of epicyclic overdensities \citep{Kuepper2009,Kuepper2011}. Such overdensities form due to the fact that tidal stripping mainly occurs near pericentric passages, leading to bursts of debris that can lead to overdensities along the stream. In particular, \citet{Ibata2020} have proposed that the density variations along the GD-1 stream are epicyclic overdensities and they purport to show that these are enough to account for the observed GD-1's density power without the need for perturbations from dark matter substructures. The effect of episodic tidal stripping and the induced epicyclic overdensities was studied in detail in \citet{Sanders2016} and \citet{Bovy2016a}, with the main take-away being that the power induced by these effects is far below that from DM substructure, essentially because in the one-dimensional projections of the density, the different episodes of stripping quickly mix together due to the velocity dispersion in the stream. 

The claim in \citet{Ibata2020} is based on a single $N$-body simulation of the  disruption of a progenitor cluster of mass $30,000\,\Msun$ for 2 Gyr, which produces a stream that somewhat matches the location and morphology of the observed GD-1 stream, but in detail looks to be much thicker, which is unsurprising, because the only way to get a long stream over only 2 Gyr is to have a massive progenitor that produces a wide stream. For such short dynamical age episodic stripping becomes important since the stripped stars don't have enough time to mix with other members. In addition to the width of the stream being in conflict with the data, such a high mass is also in conflict with the detailed collisional $N$-body simulations from \citet{Webb2018} that do reproduce the stream, in both length and width (see figures 4 and 5 in \citep{Webb2018}), where the best-fit progenitor only had an initial mass of $7,200\,\Msun$. The simulations of \citet{Webb2018} only produce overdensities because of the full disruption of the progenitor, which is the reason we exclude the part of the stream near the dissolved progenitor.  Therefore, we find \citet{Ibata2020}'s GD-1 model to not be a good representation of the observed stream and therefore their claim of not requiring dark substructures to account for the observed GD-1's density power to be unconvincing.

\section{Constraints on the dark subhalo abundance}
\label{sec:constraints}

\begin{figure}
\includegraphics[width=0.4\textwidth]{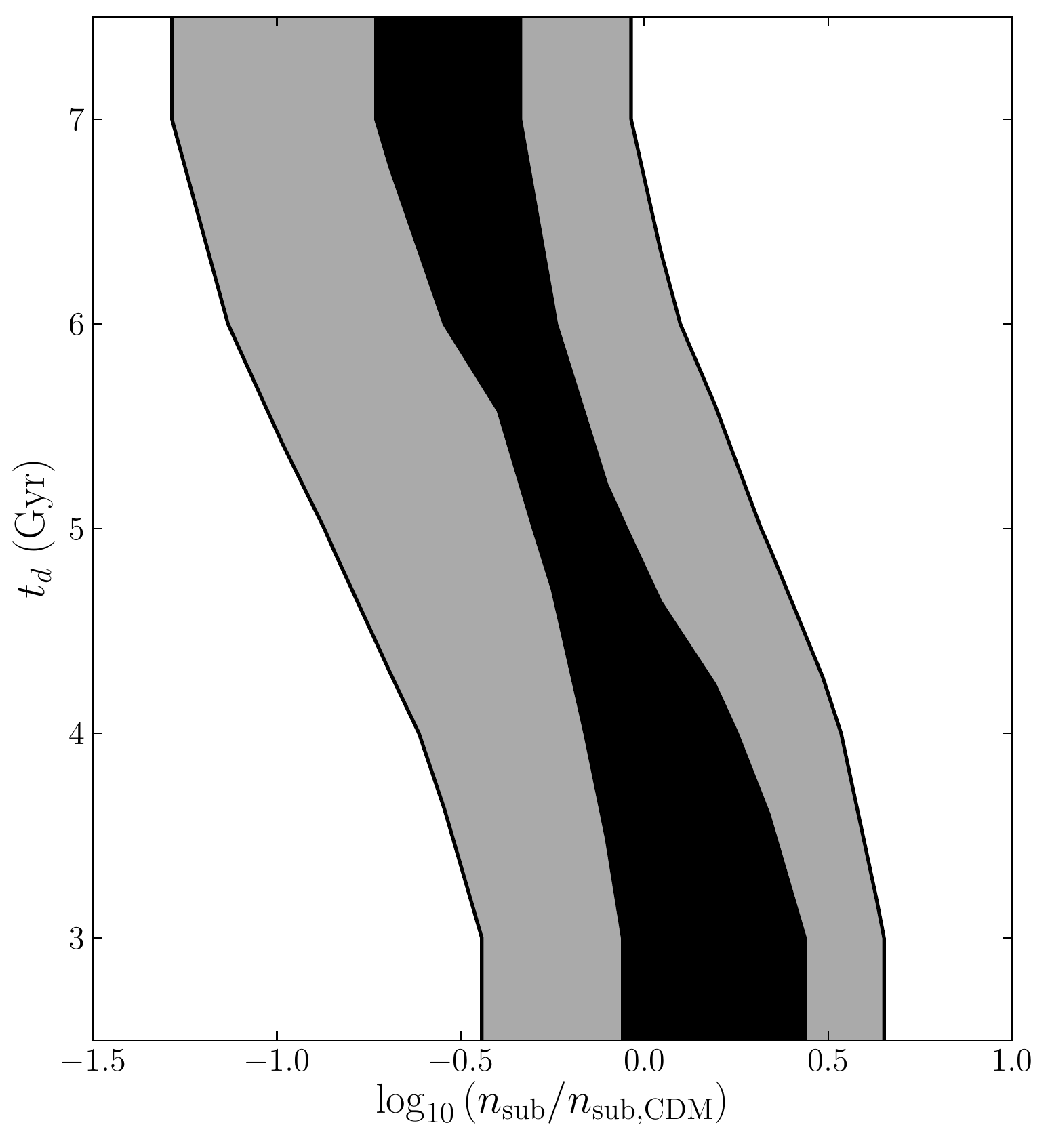}
\caption{Contour plot of the inferred relative abundance of subhalos with respect to that of the fiducial CDM case within the orbit of the GD-1 stream vs. the age of the GD-1 stream of all the accepted simulations in the ABC analysis. The contours are 1 and $2\sigma$ and show a negative correlation between the inferred subhalo abundance and age of the stream implying older stream models are consistent with lower subhalo abundance. }
\label{fig:ABC_GD1_corner_rate}
\end{figure}
\begin{figure*}
\includegraphics[width=0.95\textwidth]{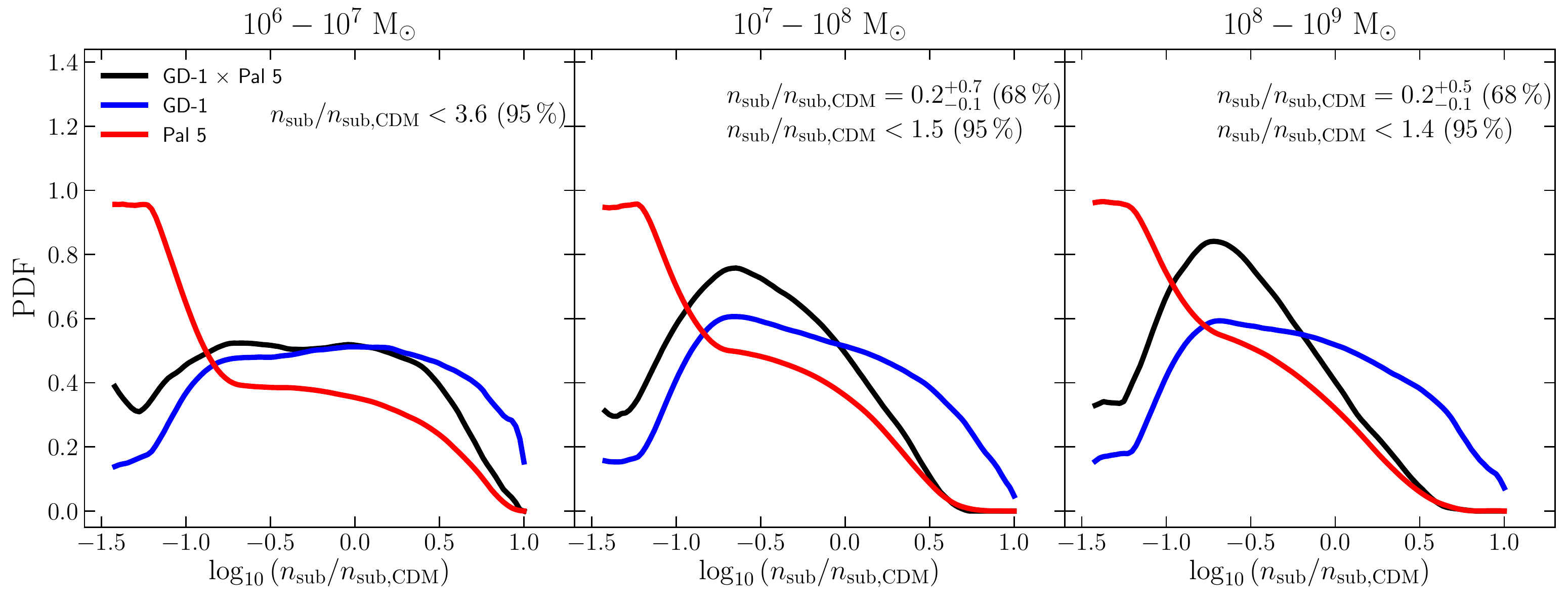}
\caption{Posterior PDFs of the subhalo abundance relative to the fiducial CDM expectation in different mass bins. The colors are same as in Figure \ref{fig:ABC_rate}. The constraints indicated are for the joint GD-1 $\times$ Pal 5 PDF.}
\label{fig:ABC_4bin}
\end{figure*}
In the previous section, we presented the methods for obtaining the stream density data of the GD-1 stream and described the mock GD-1 stream simulations and how we incorporated effects of the baryonic structures and dark matter subhalos in them. In this section, we describe how we use the GD-1 data to infer the number and mass distribution of dark subhalos in the inner Milky Way and to constrain the warm dark matter model. To improve our constraints, we use data on the Pal 5 stream previously analyzed by \citet{Bovy2016a,Banik2019} and combine it with the GD-1 data analyzed in this paper to obtain joint constraints on the dark subhalo population.

We obtain the Pal 5 stream density data from \citet{Ibata2016} and apply the same steps as in \citet{Bovy2016a} to normalize and compute its power spectrum. Following \citet{Bovy2016a} again, we generate the mock Pal 5 stream and simulate the effects of the baryonic and dark substructures on the stream as in \citet{Banik2019}.

We use the Approximate Bayesian Computation (ABC) technique to constrain different dark matter model parameters using the data and the simulations. The ABC method allows us to construct posteriors of the parameter(s) in question by comparing the simulation outputs with the data summaries, without the need for a likelihood. In our study, data summaries are the density power at different angular scales of the stream. For the GD-1 stream, we use the power at the three largest scales for both the leading and trailing arm since signal dominates noise at these scales. Likewise, for the Pal 5 stream, we use the density power at the three largest scales for its trailing arm only, because the data from \citet{Ibata2016} do not cover much of the leading arm. We choose appropriate priors for each of the dark matter model parameters that we want to constrain and run stream simulations on $\sim 10^5$ randomly drawn points from them. For the GD-1 stream, which unlike the Pal 5 stream does not have a surviving progenitor, we also marginalize over the stream age as discussed previously, while for Pal 5 we fix the age to 5 Gyr. For each realization of the mock stream density, we generate 100 more mock stream densities by adding a random Gaussian draw of the density error in the data following \citet{Bovy2016a}. This effectively gives us $10^7$ simulations. Our ABC approach accepts simulations if (a) the mock stream density power are within some pre-defined tolerance around the data summaries for both arms for the GD-1 stream and only for the trailing arm for the Pal 5 stream, and (b) the lengths of both trailing and leading arm of the mock stream (only trailing for Pal 5) are at least equal to the observed arm length. Finally, the posteriors are constructed using the accepted simulations. In order to combine the GD-1 and Pal 5 posteriors, we run simulations of both streams over the same set of points drawn from the prior and accept only those for which both Pal 5 and GD-1 density powers and lengths are accepted. In the following subsections, we present our results for the different cases of constraining the dark matter model parameters.

\subsection{Constraining the overall abundance of CDM subhalos}

For our first analysis, we constrain the abundance of subhalos in the mass range $10^{5} - 10^{9} \ \Msun$ relative to the fiducial CDM prediction. For the GD-1 stream the constraints are valid within its spherical radius of $\sim$ 20 kpc whereas for the Pal 5 stream the constraints apply within $\sim$ 14.3 kpc. Using a log-uniform prior over the range [0.03 - 10] $\times$ the fiducial CDM abundance, we run $10^6$ simulations for both GD-1 and Pal 5 stream and construct posterior PDFs using the ABC method. Figure \ref{fig:ABC_rate} shows the resulting posterior PDF for the abundance with the GD-1 stream alone (blue), the Pal 5 stream alone (red), and using the GD-1 and Pal 5 together (black). 

The GD-1 only constraint peaks at $0.7^{+ 0.9}_{- 0.5}$ at 68\% with an upper limit of $< 2.7$ at 95\% which is consistent with the fiducial CDM case. This translates to a constraint of $f_{\rm{sub}}=0.3 ^{+0.3}_{-0.2} \%$ at $68\%$ and $< 1\%$ at 95\% on the mass fraction $f_{\rm{sub}}$ in subhalos given that the total mass of the dark matter halo within 20 kpc is $\sim 10^{11} \Msun$ \citep{Bovy2013}. Very high subhalo abundances ($\gtrsim 7 \times$ CDM) are ruled out. Very low abundance ($\leq 0.05 \times$ CDM) although strongly disfavored are not completely ruled out. The latter is because we let the age of the stream vary all the way up to 7 Gyr and as evident from Figure \ref{fig:Pk_baryon_struct}, older streams accrue density power over its age due to impacts with the baryonic structures thereby reducing the amount of CDM subhalo impacts required to account for the observed density power. This is most easily seen in the 2-D histogram of inferred subhalo abundance vs. age of the stream of all the accepted simulations in Figure \ref{fig:ABC_GD1_corner_rate} which indicates that lower subhalo abundances are consistent with older stream models. The contours are $1\sigma$ and $2\sigma$ levels. If future studies are able to constrain the age of the GD-1 stream then our method can be used to put tighter constraints on the abundance of dark matter substructures within the radius of the GD-1 stream. 

The red curve in Figure \ref{fig:ABC_rate} is the posterior from the Pal 5 stream alone. It plateaus at lower subhalo abundance and puts an upper limit on the relative subhalo abundance of $< 0.9 $ at 95\% confidence. Pal 5's preference for a lower rate of subhalo impacts than that predicted by CDM is expected because it is heavily perturbed by the bar \citep{Pearson2017,Erkal2017,Banik2019}, the GMCs \citep{Amorisco2016,Banik2019}, and the spiral arms \citep{Banik2019}, so much so that the bar and the GMCs can individually account for the observed density power \citep{Banik2019}. 

Combining the GD-1 and Pal 5 posteriors yields the PDF shown by the black curve which peaks at $0.4 ^{+0.3}_{-0.2}$ at 68\% or $f_{\rm{sub}} = 0.14 ^{+0.11}_{-0.07} \%$ with an upper limit of $< 0.9$ at 95\% ($f_{\rm{sub}} \lesssim 0.3\% $), which applies within a radius of $\sim $20 kpc which encompasses both streams.

\subsection{Constraining the abundance of subhalos in different mass decades}\label{sec:massdecades}

\begin{figure}
\includegraphics[width=0.45\textwidth]{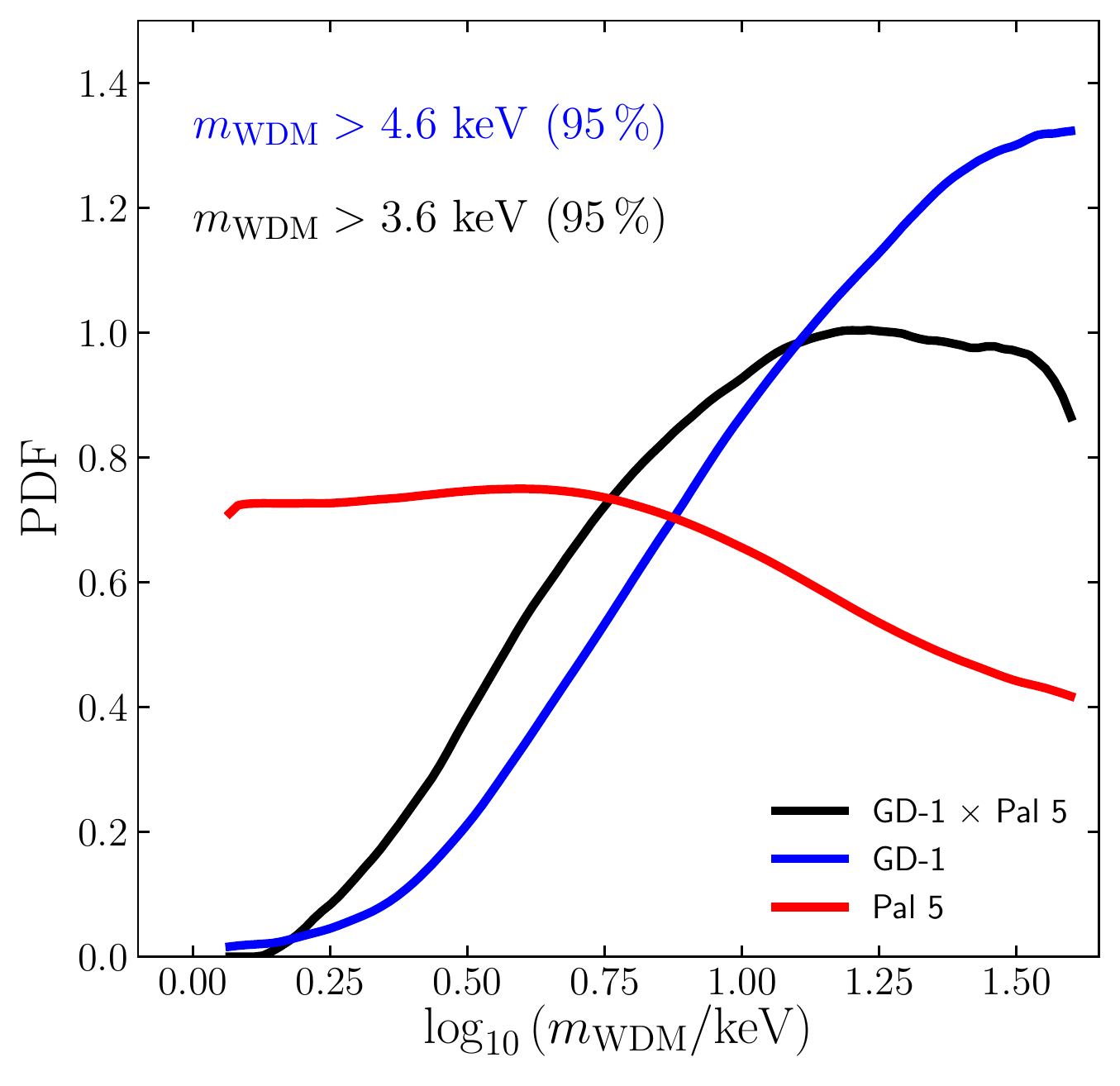}
\caption{Posterior PDF for the thermal WDM particle mass, obtained by comparing the power spectrum in the observed GD-1 (blue) and Pal 5 (red) linear density, with that arising from mock streams in presence of a population of WDM substructures. The black line shows the posterior PDF for the combined analysis of GD-1 and Pal 5 data.}
\label{fig:ABC_GD1_Pal5_mwdm}
\end{figure}

Next, we explore how the observed GD-1 and Pal 5 stream densities constrain the mass function of low-mass dark subhalos. We do this by independently varying the abundance of subhalos in the mass decades $10^{5} - 10^{6} \ \Msun$, $10^{6} - 10^{7} \ \Msun$, $10^{7} - 10^{8} \ \Msun$, and $10^{8} - 10^{9} \ \Msun$ relative to their respective fiducial CDM values. For each decade, we draw a random relative subhalo abundance from a $\log_{10}$ uniform prior on the rate relative to the CDM rate in each bin in the range $[0.03,10]$ and compute the expected number of impacts in each bin. The total number of impacts is computed by Poisson drawing from the sum of the expected number of subhalo impacts over all the mass bins. These impacts are then distributed amongst the mass bins proportional to the relative subhalo abundance in them and the stream simulations are run. Figure \ref{fig:ABC_4bin} shows the resulting posterior PDFs for the GD-1 stream in blue, Pal 5 in red, and combined GD-1 and Pal 5 in black, for the different mass decades. 

The GD-1 posterior (blue curve) is largely flat in the mass bins $10^{5} - 10^{6} \ \Msun$ (not shown here) and $10^{6} - 10^{7} \ \Msun$, implying that the level of signal in the measured power spectrum is insensitive to the abundance of subhalos in that mass bin. This is borne out of the fact that encounters with lower mass subhalos impart small scale density power which is below the level of noise in the current data. Future surveys like LSST could lower the noise level by resolving many more member stars thereby making our method sensitive to lower mass subhalos. For the higher mass bins of $10^{7} - 10^{8} \ \Msun$ and $10^{8} - 10^{9} \ \Msun$, very high subhalo abundances are less favored as indicated by the falling PDF as we approach higher abundances. Very low abundances $\lesssim 0.2 \times$ fiducial CDM are also disfavored as shown by the PDF falling sharply there. The posteriors set upper bounds of $n_{\rm{sub}}/n_{\rm{sub,CDM}}< 4.7$ and $< 4.9$ at 95\%, respectively for $10^{7} - 10^{8} \ \Msun$ and $10^{8} - 10^{9} \ \Msun$ mass bins.

\begin{figure*}
\includegraphics[width=\textwidth]{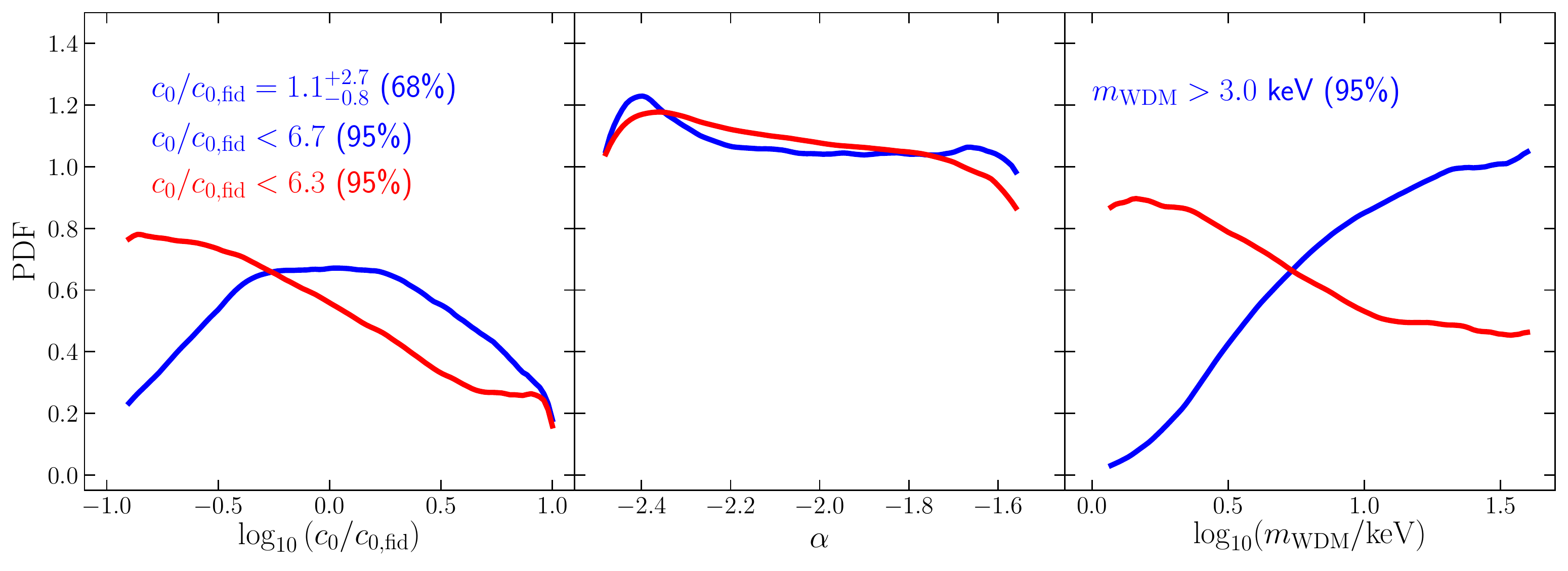}
\caption{Posterior of relative mass function amplitude $c_{0}/c_{0,\rm{fid}}$ and slope $\alpha$ and mass of the dark matter particle $m_{\rm{WDM}}$ obtained from the GD-1 stream (blue) and Pal 5 stream (red).  }
\label{fig:ABC_GD1_amp_slope_mwdm}
\end{figure*}

The Pal 5 PDF in the mass bin $10^{5} - 10^{6} \ \Msun$ stays flat and low for $\gtrsim 0.1 \times$ fiducial CDM indicating that it is unaffected by the abundance of subhalos in this mass bin. The PDF rises sharply at $\lesssim 0.1 \times$ the fiducial CDM abundance indicating its preference for very low abundance in the lowest subhalo mass bin which is also seen in all the other mass bins. This follows from the result of \citet{Banik2019} who showed that the perturbations due to the baryonic structures namely, the bar, the spiral arms and the GMCs, can account for Pal 5's observed density power. Therefore, a very low subhalo abundance which results in no significant effects on the stream is preferred. For the bin $10^{6} - 10^{7} \ \Msun$, the PDF falls to 0 sharply at $\sim  10\times$ CDM abundance indicating abundances higher than that are ruled out, while placing an upper bound of $< 3 \times$ CDM at 95\% level. For the upper two bins, abundances $\gtrsim 3\times$ CDM are ruled out, while placing upper bounds of $<1.6 \times$ and $< 1.3 \times$ CDM at 95\% confidence. 

The combined PDF is flat over the range of the prior in the lowest subhalo mass bin and hence does not constrain its abundance (for this reason we do not show it). For the mass bin $10^{6} - 10^{7} \ \Msun$, the combined PDF falls sharply at $\gtrsim 3 \times$ fiducial CDM abundance, placing an upper bound of $< 3.6 \times$ fiducial CDM at 95\% confidence. For the mass bins $10^{7} - 10^{8} \ \Msun$ and $10^{8} - 10^{9} \ \Msun$, the posterior peaks at relative abundances of $0.2 ^{+0.7}_{-0.1}$ and $0.2^{+0.5}_{-0.1}$ at 68\% respectively. At 95\% confidence the upper bounds are $<1.5$ and $<1.4$, respectively.

\begin{figure}
\includegraphics[width=0.45\textwidth]{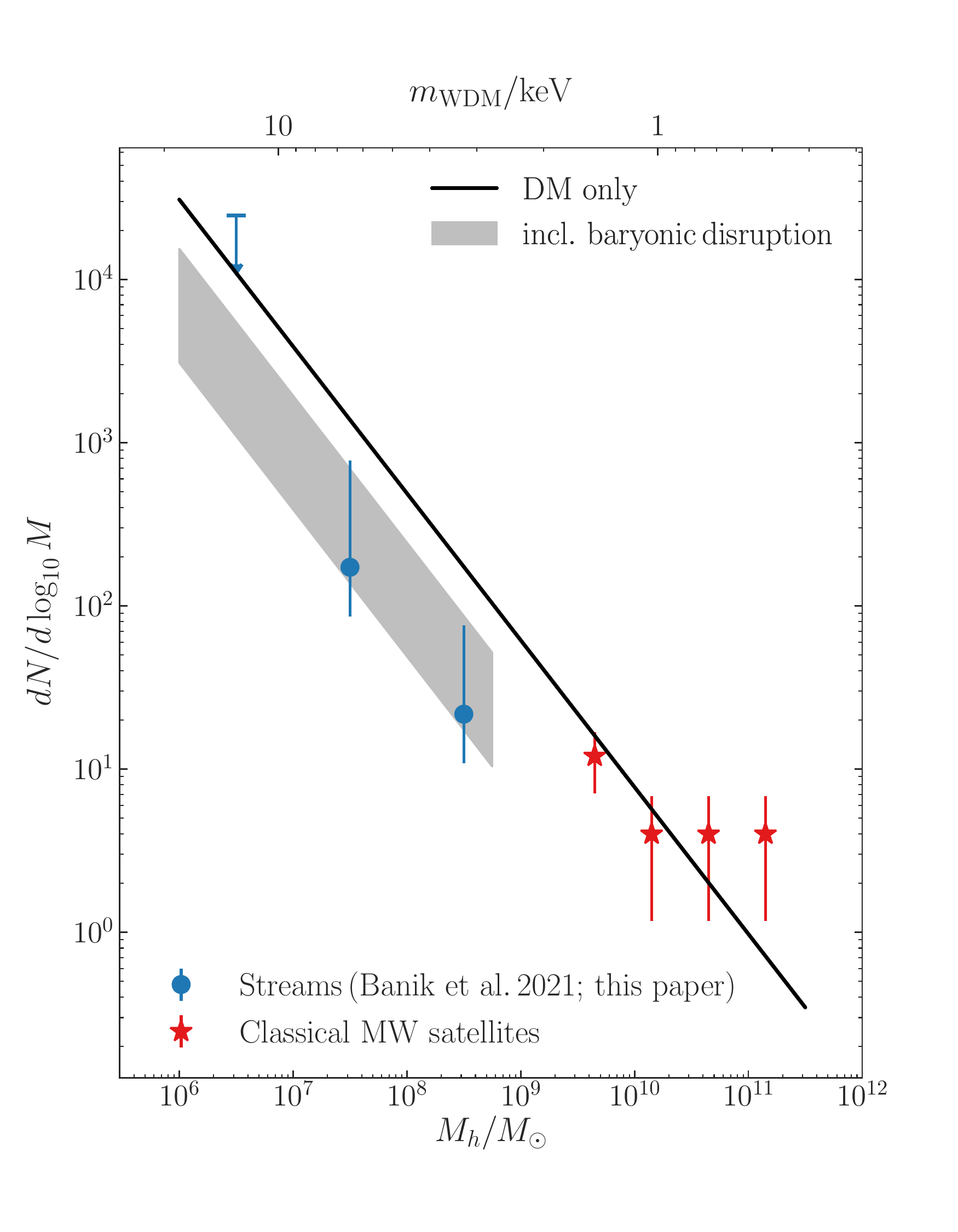}
\caption{Measuring the subhalo mass function using the combined statistics of Pal 5 and GD-1 streams. The black solid line indicate the subhalo mass function in the fiducial CDM only model. The gray shaded region shows mass function as a result subhalo disruptions by a factor of 2 to 10, due to the disk and/or the Milky Way potential. The blue errorbars represent our stream measurements and is consistent with fiducial mass function taking into account subhalo disruption.    }
\label{fig:subhaloMF_4bin}
\end{figure}

\section{Constraints on WDM}
\label{sec:mwdm}

\subsection{Mass of the dark matter particle}

Next, we use the stream data to constrain the mass of the dark matter particle $m_{\rm{WDM}}$, considering dark matter is composed entirely of thermal relic dark matter (refer to section \ref{sec:subhalos} on methods of incorporating thermal relic WDM subhalos in our stream simulations). Since $m_{\rm{WDM}}$ is a parameter with units whose magnitude is unknown, we consider a uniform prior in $\log_{10}(m_{\rm{WDM}})$ in the range [1-50] keV, which ensures that our prior is sufficiently non-informative (we consider other priors in Sec.~\ref{sec:wdmwithclassical} below). The upper bound of 50 keV corresponds to a half-mode mass of $\sim 4\times 10^{4} \ \Msun$ which is well below the sensitivity of our stream analysis method or any other current method for constraining the abundance of subhalos. For each drawn $m_{\rm{WDM}}$ from the prior, we run the stream simulation as described in section \ref{sec:subhalos} and then compare the resulting stream density power spectrum with that of the observed power within the ABC framework. 

The resulting posterior PDFs are shown in Figure \ref{fig:ABC_GD1_Pal5_mwdm}. The GD-1 posterior, shown by the blue curve, puts a lower limit of $m_{\rm{WDM}}> 4.6$ keV at 95\%. Warm dark matter models with particle mass $< 1.5$ keV are ruled out as seen by the PDF dropping to $\sim 0$. The PDF plateaus for $m_{\rm{WDM}} \gtrsim 31$ keV indicating masses above that are equally preferred by the GD-1 stream. 

The Pal 5 posterior, shown by the red curve, does not constrain $m_{\rm{WDM}}$ but it prefers a low $m_{\rm{WDM}}$. The PDF has a considerably high value of $\sim 0.4$ for a dark matter particle mass $\gtrsim 39 \ (\sim 10^{1.6})$ keV, for which the abundance of subhalos in the mass range $10^{5} - 10^{9} \ \Msun$ is similar to that in the fiducial CDM case. This seems counterintuitive at first since the density perturbation imparted on Pal 5 by the effect of the bar, spiral arms and the GMCs in our Galaxy is similar to that by the subhalos in the fiducial CDM case (see Figs. 15 and 17 in \citet{Banik2019}), and can account for the observed density power. The reason behind this is that in running the stream simulations we marginalize over the uncertain nature of interaction between the GMCs and the stream arising from the short life span of GMCs relative to that of the dynamical age of the stream, shown in Figure 15 in \citet{Banik2019} by the gray shaded region. As evident from that figure, there will be many realizations in which the density power due to the GMCs will be an order of magnitude lower than the median power. Such cases will require a CDM like subhalo abundance (higher mass $m_{\rm{WDM}}$) to account for the observed power. This behavior of the Pal 5 PDF is also evident in Figure \ref{fig:ABC_rate} where it has a considerable value at CDM-like abundances. 

The combined posterior, shown by the black curve, puts a constraint of $m_{\rm{WDM}}> 3.6$ keV at 95\% and rules out particle masses $<1.3$ keV. It is important to point out that the upper bound of the prior is somewhat arbitrary and, in particular, that one can pick an arbitrarily large upper bound of the prior and obtain a higher constraint on the $m_{\rm{WDM}}$. However, we choose 50 keV since a thermal relic dark matter candidate of that mass will correspond to a half-mode mass which is well below the sensitivity of our stream studies. Because the joint PDF declines slightly at higher $m_{\rm{WDM}}$, because of Pal 5's preference for low $m_{\rm{WDM}}$, cutting at posterior 38.8 keV yields $> 4.4$ keV at 95\%, similar to the constraint from GD-1 alone.

\subsection{Constraints on the amplitude, slope, and mass of the dark matter particle}

Next, we explore the constraints on the mass of the dark matter particle if the amplitude and slope of the mass function are set free to vary. Referring back to the notations used in equations~\eqref{eq:dndMc} and \eqref{eq:wdmcdm}, we consider a $\log_{10}$ uniform prior on the relative amplitude $c_{0}/c_{0,\rm{fid}}$ in the range $[1/10,10]$, a uniform prior on the slope $\alpha$ in the range $[-2.5,-1.5]$ and the same $\log_{10}$ uniform prior on $m_{\rm{WDM}}$ in the range $[1,50]$~keV. The posteriors are shown in Figure \ref{fig:ABC_GD1_amp_slope_mwdm}, where the GD-1 stream PDFs are shown in blue and the Pal 5 stream PDFs in red. 

The GD-1 stream posterior PDF for the relative amplitude has a peak at $c_{0}/c_{0,\rm{fid}}=1.1^{+2.7}_{-0.8}$ at 68\% confidence and puts an upper limit of 6.7 at 95\%. The posterior on the slope $\alpha$ is largely flat with a preference for lower slopes. 
Finally, the posterior on $m_{\rm{WDM}}$ is $> 3$ keV at 95\%. This constraint on $m_{\rm{WDM}}$ is weaker than the previous analysis where the amplitude and slope of the mass function were held fixed at their respective fiducial value. This is because the amplitude is negatively correlated with $m_{\rm{WDM}}$ and therefore lower values of $m_{\rm{WDM}}$ can fit the data if the amplitude is high. 

For Pal 5, the posterior of the relative amplitude plateaus at a lower value and puts a constraint of $< 6.3$ at 95\% confidence. Similar to GD-1, the posterior on the slope is largely flat with preference towards lower values and insensitive to the stream statistics. The posterior on the dark matter particle mass prefers a low value which is due to the same reason as explained in the previous section.   

Since the individual GD-1 and Pal 5 PDFs for $m_{\rm{WDM}}$ are quite different, obtaining a sufficient number of accepted samples using the combined data is difficult. We do not pursue the joint analysis for this more general model because GD-1 gives a stronger constraint than the combined analysis in the previous sub-section where we only varied $m_{\rm{WDM}}$. However, in the next sub-section we include both the GD-1 and Pal 5 stream constraints into a more general WDM analysis that also includes the classical Milky Way satellites.

\subsection{Including classical Milky Way satellites.}\label{sec:wdmwithclassical} 

We have so far performed a conservative analysis of the constraints on the mass function of dark matter subhalos arising from the study of density variations in Milky-Way streams only. We explore now the consequences of also taking into account the measurement of the high-mass end of the mass function using observations of classical Milky Way satellites. For this, we use the compilation of properties of the classical satellites (with stellar mass $M_* > 10^5\,\Msun$) within 300 kpc from \citet{GarrisonKimmel19a} (these are the LMC, SMC, Sagittarius, Fornax, Sculptor, Leo I and II, Sextans, Ursa Minor, Carina, Draco, and Canes Venatici I). We assign dark-halo masses for all of these satellites using the stellar-mass vs. halo-mass relation for satellites given in \citet{2017ARA&A..55..343B}, obtained from abundance matching,
\begin{equation}
    \log_{10} \left(\frac{M_h}{10^{11}\,\Msun}\right) = 0.468\,\log_{10}\left(\frac{ M_*}{3\times 10^8\,\Msun}\right)\,.
\end{equation}
We then compute the classical-satellite mass function by counting the number of satellites in $\log_{10}M_h/\Msun$ bins of width 0.5 between the lower limit in stellar mass of $9.4$ and $11.4$; the uncertainty on these numbers is Poisson distributed. The dark-matter subhalos probed by our stream measurements live within $\approx20\,\mathrm{kpc}$ from the Galactic center. To be able to combine these stream measurements with the mass function derived from the classical satellites, we extrapolate their abundance assuming that the radial distribution of subhalos follows the Einasto profile from Eqn.~\eqref{eq:dndMc}. We do this for the measurements of the subhalo abundance in different mass decades from section~\ref{sec:massdecades} (for the lowest mass bin of $10^{6} - 10^{7} \Msun$, we show the 95\% upper limit, because of the lack of a peak in the PDF for that bin). The resulting subhalo mass function is shown in Figure \ref{fig:subhaloMF_4bin}.

We compare the observed subhalo mass function in the Milky Way in Figure \ref{fig:subhaloMF_4bin} to the predictions from dark-matter-only (DM-only) CDM simulations (black line), the mass function part of Eqn.~\eqref{eq:dndMc}, multiplied by a factor of 1.6 determined by fitting the observed mass function (see below). While the classical satellites are typically found at great distance from the Galactic disk and their abundance is therefore not expected to be strongly affected by subhalo disruption due to the disk, the subhalo abundance probed by our stream measurements in the inner Milky Way is likely reduced with respect to the DM-only prediction. A range of plausible reduction factors between 10\% and 50\% is indicated by the gray band in Figure \ref{fig:subhaloMF_4bin}. It is clear that our measurements of the abundance of low-mass dark-matter subhalos is in good agreement with the predictions from CDM, including the effect of baryonic disruption.

To obtain further constraints on the mass of WDM from the combined set of measurements from classical satellites and streams, we fit the data in Figure \ref{fig:subhaloMF_4bin} with models for the WDM mass function from Eqn.~\eqref{eq:wdmcdm}. Specifically, we vary the mass of WDM, $m_{\rm{WDM}}$, the logarithmic slope $\alpha$ and the normalization $c_0$ of the CDM part of the mass function (see Eqn.~[\ref{eq:dndMc}]), and the fraction of subhalos $f_\mathrm{survive}$ in the inner Milky Way that survives tidal disruption by the disk, while fixing all other parameters related to the radial profile or the WDM mass function. We use  logarithmic priors on $c_0$ (between 0.01 and 100 times the fiducial value given below Eqn.~[\ref{eq:dndMc}]) and on $f_\mathrm{survive}$ (conservatively between 0.1\% and 50\%). For $\alpha$, our standard prior is flat between $-1.95$ and $-1.85$, the range found in numerical simulations of halo formation \citep{Springel2008}, but we also investigate the effect of a looser uniform prior between $-3$ and $-1$. For $m_{\rm{WDM}}$, we explore a variety of priors to assess the impact of the prior. 

The likelihood that we use in the fit is composed of two parts, that of the classical satellite counts and that of our new stream constraints on the subhalo abundance. For the abundance of the classical satellites, we compute the number of classical satellites that would exist in each model and compare it to the observed number using the Poisson distribution. For the subhalo abundance measurements from streams, we use the PDFs from the determinations of the subhalo abundance in different mass decades from section~\ref{sec:massdecades}. We only use those from the two highest-mass bins, because the upper limit in the $10^6$ to $10^7\,\Msun$ bin is too weak to be useful for the WDM constraint. To be able to easily use these PDFs, we approximate the curves in Figure~\ref{fig:ABC_4bin} with smooth functions: for $10^7$ to $10^8\,\Msun$
\begin{align}
    \ln p(r=\log_{10}[n_\mathrm{sub}/n_{\mathrm{sub, CDM}}]) = & -\frac{|r+0.5|^{2.5}}{2\times 0.5^2}\,,
\end{align}
and for the skewed PDF for $10^8$ to $10^9\,\Msun$
\begin{align}
    \ln p(r=& \log_{10}[n_\mathrm{sub}/n_{\mathrm{sub, CDM}}]) \nonumber\\ 
    & = -\frac{|r+0.7|^{2}}{2\times 0.3^2}\qquad (r < -0.7)\\
    & = -\frac{|r+0.7|^{2}}{2\times 0.6^2}\qquad (r \geq -0.7)\,.
\end{align}
For each model, we compute $\log_{10}[n_\mathrm{sub}/n_{\mathrm{sub, CDM}}]$ in each mass bin and use these probabilities to compute the likelihood. We use the affine-invariant \emph{emcee} sampler to run MCMC analyses using this likelihood and priors \citep{Goodman10a,emcee}.

To avoid the necessity of a hard cut-off in the prior on $m_{\rm{WDM}}$ as above, for a first analysis we use a conservative prior that is flat in $1/m_{\rm{WDM}}$. With this prior, we find a $95\%$ lower limit of $m_{\rm{WDM}} > 4.9\,\mathrm{keV}$. While the posterior on the logarithmic slope of the subhalo mass function $\alpha$ is the same as the prior, we do constrain the amplitude of the mass function to be $1.6^{+0.6}_{-0.5}$ times the fiducial value given below Eqn.~\eqref{eq:dndMc} and the fraction of subhalos in the inner Milky Way that survives disruption by the disk to be $f_\mathrm{survive} = 0.21^{+0.17}_{-0.11}$; although for the latter the PDF is wide, skewed, and peaks at $f_\mathrm{survive} \approx 0.1$. This indicates that a large fraction of dark-matter subhalos is disrupted by the disk in the inner Galaxy. Relaxing the prior on $\alpha$ to be between $-3$ and $-1$, we find that $\alpha = -2.1\pm0.3$; however, the PDFs on all other parameters are much wider, because when allowing such extreme values of $\alpha$ odd fits to the data become possible (e.g., fits with low $m_{\rm{WDM}}$, high mass-function amplitude, and small $\alpha$ are able to fit the data, but such values are not supported by numerical simulations of halo formation). For the purpose of constraining $m_{\rm{WDM}}$, we therefore follow the results from numerical simulations in setting $\alpha \in [-1.95,-1.85]$. 

A less conservative prior on $m_{\rm{WDM}}$ is a flat prior on $\log m_{\rm{WDM}}$ as we have used above; equivalently, we can assume a flat prior on $\log_{10} M_{\mathrm{hm}}$, the half-mode mass, which makes our results easier to compare to the best current constraints on $m_{\rm{WDM}}$ from strong lensing (see \citet{Gilman19a}). To be directly comparable to the strong lensing results, we use a uniform prior on $\log_{10} M_{\mathrm{hm}}/\Msun \in [4.8,10]$. In this case, we find a 95\% lower limit of $m_{\rm{WDM}} > 6.3\,\mathrm{keV}$ or, equivalently, $M_{\mathrm{hm}} < 4.3\times 10^7\,\Msun$. The PDFs for $\alpha$, the normalization of the mass function, and $f_{\mathrm{survive}}$ are similar as those for the flat prior on $1/m_{\rm{WDM}}$. Compared to the results from strong lensing, our constraint on $m_{\rm{WDM}}$ is a keV stronger.

\section{Discussion and Conclusions}
\label{sec:conclusions}

In this paper we used the density power spectrum of the linear stellar density of the GD-1 stream, whose data was obtained from \textit{Gaia} and PanSTARRS, to infer the abundance of dark matter substructure within 20 kpc of the Galactic center. Assuming that the underdense region centered at $\sim -40^{\circ}$ was caused due to the stream progenitor disruption $\sim 500$ Myr years ago, we constructed GD-1 models of dynamical age 3,4,5,6 and 7 Gyr. We studied the cumulative effects of the known baryonic structures namely, the bar, spiral arms, the GMCs and the GCs on the stream density by computing its density power spectrum and showed that it is insufficient to account for the observed level of power in the stream data. Including dark matter substructures however, accounts for the observed density power. 

In analyzing the GD-1 stream we have ignored the \textit{spur} and \textit{blob} structures which were found in \citet{Price-Whelan18a}. Recently, \citet{Bonaca2018} showed that the spur feature could be due to an encounter with a compact substructure with a mass in the range $10^6 - 10^8 \ \Msun$ and of scale size $\lesssim 10$ pc. Based on the relatively large ($\gtrsim 10$ kpc) separation between the GD-1 gap and the known baryonic objects (GMCs, GCs and known Milky Way satellites) over the last 1 Gyr, they hypothesized that the likely candidate which might have caused the spur is a dark matter subhalo. 

Such a subhalo would be far more dense than standard CDM substructure: a standard CDM subhalo of even the lower end of their mass range, $10^6 \ \Msun$, has a scale radius of $\sim 100$ pc in our model and is therefore totally incompatible with their requirements of a spur-inducing mass. Using an extensive set of collisional N-body simulations of the formation of the GD-1 stream, \citet{Webb2018} found that the impact time preferred by \citet{Bonaca2018}, $\sim 500$ Myr ago, is a likely time at which the progenitor fully disrupted during its last pericentric disk passage, creating the density gap at $\phi_1 \approx -40^\circ$; the spur may then have formed in the process of the final disruption. \citet{Boer2019} studied the effect of classical satellites on GD-1 and found that the Sagittarius dwarf can create a spur during a close encounter with GD-1. It is worth emphasizing that the 5 Sgr models which showed observable density fluctuations were hand picked out 1000 Sgr models constructed by sampling its current phase space uncertainty to create big effect. As they have correctly pointed out, this is therefore an unlikely possibility and not a generic prediction, as shown by our Monte Carlo analysis of all of the dwarf satellites discussed in section \ref{sec:GMC_GC}, which showed that there is only a very small amount of density variations expected from the influence of all classical satellites, including Sgr. Regardless of what causes the spur and blob features, they occur far enough from the main stream track that they are not included in our data and our CDM/WDM modeling is therefore unaffected by them. 

We have also ignored the diffuse envelope of stars around the stream \citep{Carlberg2018} as found around GD-1 in \citet{Malhan2019} where it was claimed that the progenitor globular cluster of the GD-1 stream was accreted along with its host dwarf galaxy into the Milky Way halo and the diffuse envelope of stars is what remains of the tidally disrupted dwarf galaxy. In this scenario, the GD-1 stream is then the same thin stream as in our model cocooned within this diffuse field of stars that has no bearing on our analysis and can be excluded like the background stars. \citet{Carlberg2018} also found that in approximate $N$-body simulations of stellar-stream formation in an evolving dark-matter halo with small-scale CDM substructure removed, streams displayed significant density variations with almost the same power as that from CDM substructure, although they did not identify a physical reason for this effect. Conservatively, if this effect contributes to the power that we observe, our resulting subhalo abundance would drop by a factor of two---still within the range of abundances predicted by hydrodynamical simulations--because it would mean that we can explain half of the observed power without small-scale CDM substructure (and power scales approximately linearly with subhalo abundance; see \citet{Bovy2016a}). However, more work identifying the reason for the density variations in the evolving-DM-halo simulations is required to be able to fully assess the impact of these effects.

We included the Pal 5 stream data from CFHT and applied Bayesian statistics in the form of Approximate Bayesian Computation technique to constrain the dark matter subhalo abundance in the mass range $10^{5} - 10^{9} \ \Msun$ relative to the fiducial CDM subhalo abundance. We found that GD-1 alone prefers a subhalo abundance that is consistent with the CDM prediction. However, because Pal 5 is heavily perturbed by the baryonic structures, it favors a low abundance of subhalos. A joint analysis of the GD-1 and Pal 5 data sets infers a total abundance of dark subhalos, normalised to standard CDM predictions using dark-matter-only simulations, of $n_{\rm sub}/n_{\rm sub, CDM} = 0.4 ^{+0.3}_{-0.2}$. Alternatively, our result can be expressed as the fraction $f_{\rm{sub}}$ of the dark-matter halo within 20 kpc that is in bound substructures: $f_{\rm{sub}} = 0.14 ^{+0.11}_{-0.07} \%$. This number is fully consistent with the depletion of a CDM population of subhalos by tidal disruption due to the massive Galactic disk. 

We also explored how the stream statistics can be used to constrain the subhalo abundances in different mass decades. We found that the stream statistics is insensitive to the abundance of subhalos below $10^{6} \ \Msun$. With the current level of noise we obtained measurements of the subhalo abundance in the mass bins $10^{7} - 10^{8} \ \Msun$ and $10^{8} - 10^{9} \ \Msun$ that are both $n_{\rm sub}/n_{\rm sub, CDM} \approx 0.2$, and an upper limit in the mass bin $10^{6} - 10^{7} \ \Msun$ of $n_{\rm sub}/n_{\rm sub, CDM} < 3.6$ at 95\,\% confidence, using both the Pal 5 and GD-1 streams. Thus, for the first time, we measure that the dark-matter subhalo abundance down to $10^{7} \,\Msun$ is consistent with a CDM population of subhalos depleted by tidal disruption due to the massive Galactic disk. Future surveys such as LSST and WFIRST will resolve more stars along the streams with better precision thereby lowering the level of noise in the data and making our analysis sensitive to density variations at smaller angular scales and hence lower mass subhalos. 

While the results from GD-1 and Pal 5 are consistent, a true discrepancy could result from a scenario where the Pal 5 stream that we see is a remnant of a longer, older stream that got disrupted by the combined effects of the baryonic structures and a CDM-like population of subhalos. This speculation is based on what we found in our Pal 5 stream simulations where we found that including the baryonic structures and subhalos of CDM abundance resulted in disrupting and truncating the stream. This could in principle be verified by means of a chemical tagging analysis of the Pal 5 stream stars and the background field of stars but given that the stream members are extremely faint it will be very difficult to segregate them from the background stars. So far, we have only considered Pal 5's trailing arm, because the leading arm is outside the footprint of the CFHT survey. Recently, \citet{Starkman2019} found $\sim 7^{\circ}$ of stream along the leading arm of Pal 5 from \textit{Gaia}. Future spectroscopic follow up of those stars could allow us to dynamically model the leading arm and include it in our study which would improve the constraints coming from Pal 5 and could sharpen or ameliorate the discrepancy of Pal 5's preferred subhalo abundance with that inferred from the GD-1 stream.  

Finally, we investigated how the stream data can be used to constrain the mass of the dark matter particle in a model where the entire population of dark matter is a thermal relic from the early Universe. These results are discussed in more detail in a companion paper to this \citep{Banik2020L}, but we summarize the results here for the sake of completeness. First, we obtained a lower limit on $m_{\rm{WDM}}$ using the GD-1 data while keeping all the parameters of the mass function fixed at their fiducial values, finding a best lower limit of $m_{\rm{WDM}}> 4.6$ keV at 95\%. Next, we explored how these constraints change if we let the amplitude and slope of the mass function vary within reasonable prior ranges, which weakens the constraint to $m_{\rm{WDM}} > 3$ keV at 95\%. Finally, we used the subhalo abundance measurements in the mass bins $10^{7} - 10^{8} \ \Msun$ and $10^{8} - 10^{9} \ \Msun$ that we obtained in section \ref{sec:massdecades} from the joint GD-1 and Pal 5 analysis, and combined them with the classical Milky Way satellite counts in the mass range $10^{9.4} - 10^{11.4} \ \Msun$ to constrain $m_{\rm{WDM}}$. Using all of these data, we find a 95\% lower limit of $m_{\rm{WDM}} > 6.3\,\mathrm{keV}$.

{\bf Acknowledgements.---} NB acknowledges the support of the D-ITP consortium, a programme of the Netherlands Organization for Scientific Research (NWO) that is funded by the Dutch Ministry of Education, Culture and Science (OCW). JB received support from the Natural Sciences and Engineering Research Council of Canada (NSERC; funding reference number RGPIN-2015-05235). Portions of this research were conducted with high performance research computing resources provided by Texas A\&M University (\url{https://hprc.tamu.edu}).

{\bf Data availability.---} No new data were generated in support of this research. The GD-1 density data is from \citet{Boer2019} and the Pal 5 density data is from \citet{Ibata2016}. Both datasets are publicly available in those articles. 

\bibliographystyle{mnras}
\bibliography{analyzeGD1}{}

\appendix

\section{Effect of the size of cut around the progenitor}
\label{sec:cut_width}

In this appendix we demonstrate how the size of the cut around the progenitor affects the density power spectrum of the stream. The details of the progenitor's disruption affect the stream density mainly within a few degrees of the progenitor \citep{Bovy2016a}, but to be conservative we chose a cut of $12^{\circ}$ around the supposed location of the progenitor for all the analyses in this paper. In Figure \ref{fig:cut_width}, we show how the power spectrum of the GD-1 stream goes up as the size of the cut is reduced from the fiducial case of $12^{\circ}$ (in green) to $8^{\circ}$ (in red) and to $4^{\circ}$ (in black). This happens because as we remove more stream we remove density fluctuations making the overall stream smoother and shorter. The error bars represent the power in the data and the solid lines represent the power in mock GD-1 streams analyzed with the same cut that are 7 Gyr old and had encounters with all the previously discussed baryonic structures and fiducial CDM abundance of subhalos. The gray shaded region in figure is the $2\,\sigma$ dispersion of power of the mock GD-1 stream with the $4^{\circ}$ cut. 

\section{Scenario in which the GD-1 progenitor disruption led to the gap at $\phi_{1} = -20^{\circ}$ }
\label{sec:prog-20}

\begin{figure*}
\includegraphics[width=0.9\textwidth]{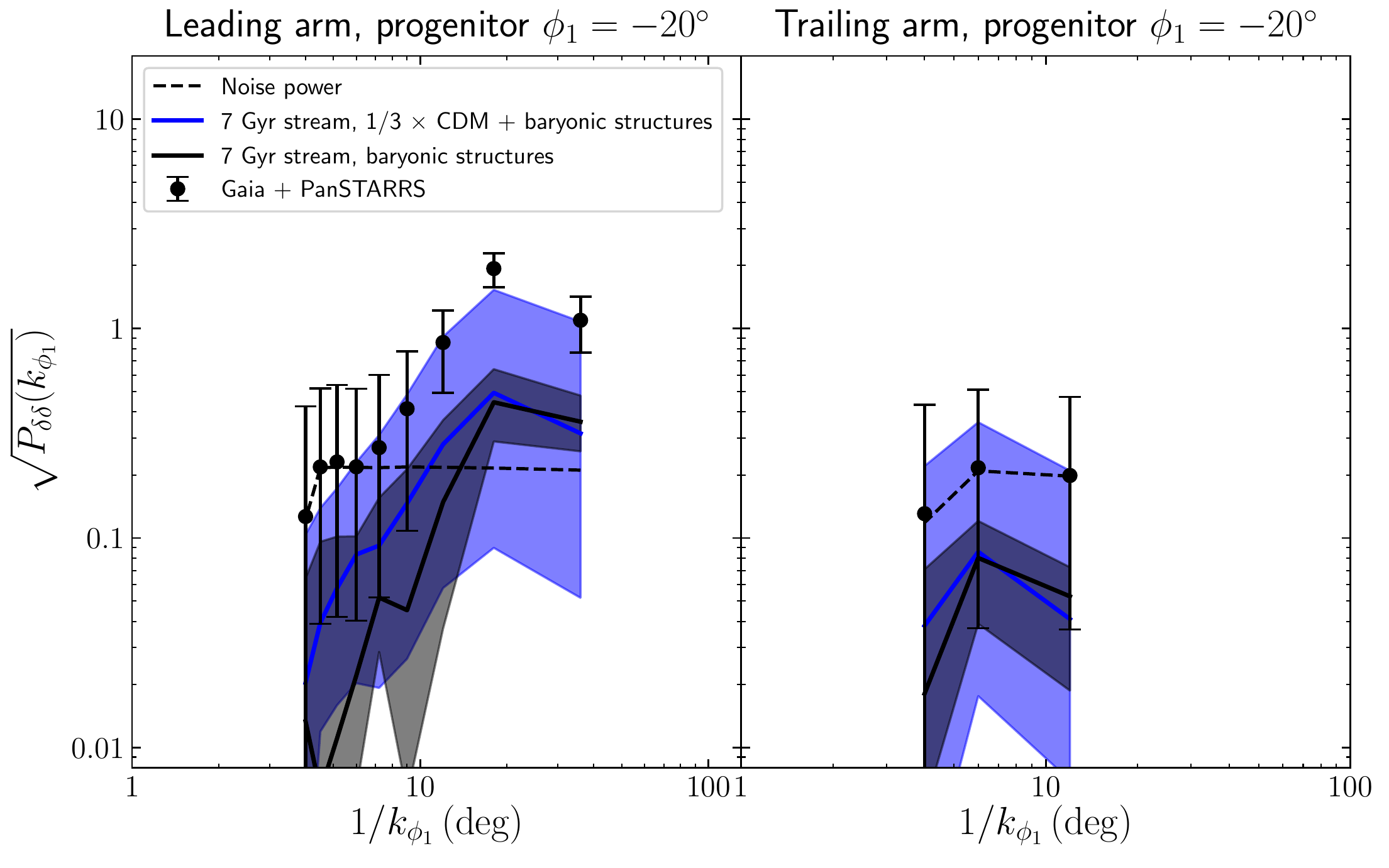}
\caption{Density power spectrum of the GD-1 stream assuming that its progenitor disruption resulted in the gap at $\phi_{1} = -20^{\circ}$. The left panel shows the power spectrum of the leading arm and the right panel shows the power spectrum of the trailing arm. The black errorbars represent the density power in the data and the blue solid line shows the median density power of 1000 mock realizations of a 7 Gyr old stream that had impacts from the baryonic structures and one-third of the fiducial populations of CDM subhalos. The blue shaded region shows the $2\,\sigma$ dispersion of power of these realizations. The black line and the gray shaded region is the median power and the $2\,\sigma$ dispersion of density power in the absence of dark matter subhalo impacts.}
\label{fig:prog-20}
\end{figure*}

In this appendix, we investigate how the results of this paper are affected had we assumed the disruption of the progenitor of the GD-1 stream led to the gap at $\phi_{1} = -20^{\circ}$. As in the fiducial case, we consider GD-1 stream models of dynamical age 3, 4, 5, 6 and 7 Gyr and follow the same steps to generate these models and incorporate effects from the baryonic structures and dark matter subhalos. As before, we cut out $12^{\circ}$ around the progenitor to exclude effects from the disruption of the progenitor which makes the trailing arm $10^{\circ}$ and the leading arm $34^{\circ}$ along $\phi_{1}$. We find that just like in the fiducial GD-1 stream model, in the absence of dark matter substructures, the density power of the mock stream can not account for the observed power. This is shown in Figure \ref{fig:prog-20} where the power spectrum of the observed GD-1 stream (indicated by error bars) is compared with that of the 7 Gyr old GD-1 model which was evolved in the presence (blue line and blue shaded region) and absence (black line and gray shaded region) of a fiducial CDM population of dark matter substructures. The density power of the observed trailing arm is at the level of noise power as most of the density information was removed by the cut around the progenitor, rendering the power spectrum of this arm ineffective for inferring the properties of the perturbers. We also find the density power of younger stream models to be consistently lower than older stream models similar to the trend seen in Figure \ref{fig:Pk_CDM_baryonicstruct}. Comparing Figure \ref{fig:prog-20} with Figure \ref{fig:Pk_CDM_baryonicstruct}, it can be seen that this alternate GD-1 model requires a somewhat higher CDM subhalo abundance to be consistent with data.

\begin{figure}
\includegraphics[width=0.4\textwidth]{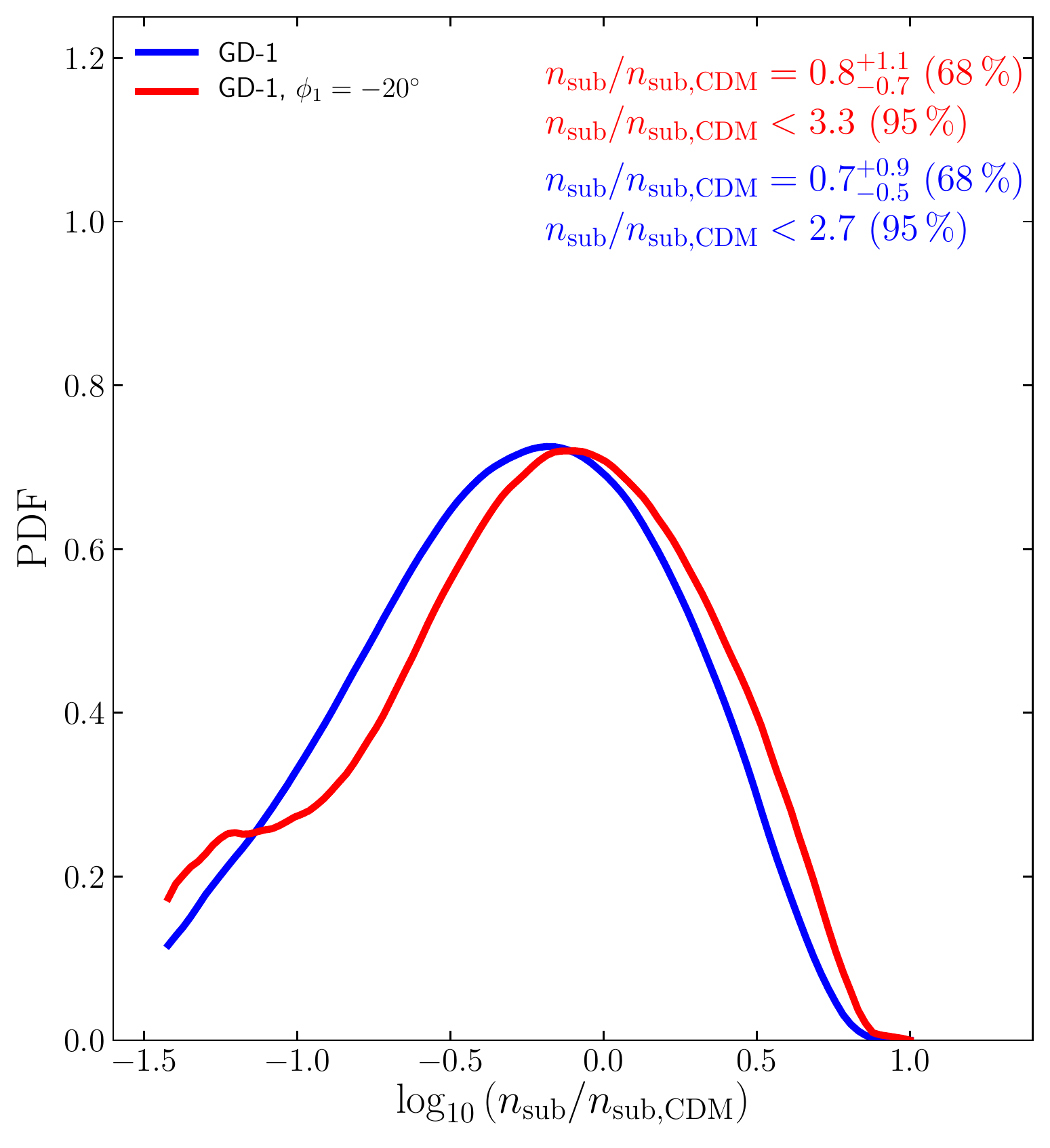}
\caption{Posterior distribution and the predicted abundance in this alternate GD-1 model (red) and in the fiducial case (blue).}
\label{fig:prog-20_ABC_rate}
\end{figure}

\begin{figure*}
\includegraphics[width=0.8\textwidth]{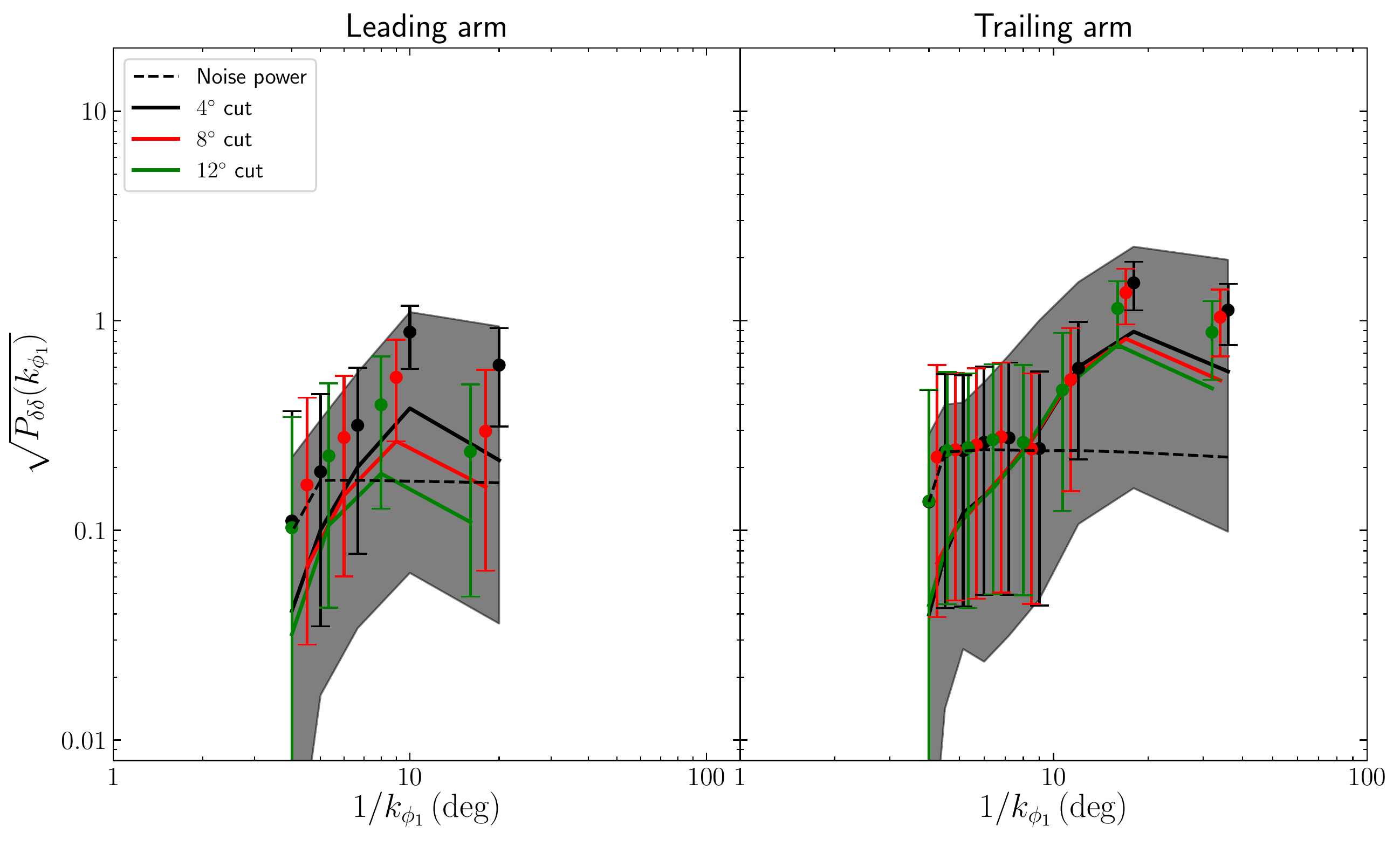}
\caption{Density power spectrum of the leading and trailing arms of the GD-1 stream for different choices of how much stream near the progenitor is excised from the analysis. The points with the errorbars represent the power in the stream data with different extents of the stream around the progenitor cut out. The dashed horizontal line represent the noise power in the data for the $4^{\circ}$ cut. The solid lines represent the density power in mock GD-1 stream models of 7 Gyr age as a result of its gravitational encounters with baryonic structures and sets of CDM subhalos. The $2\,\sigma$ dispersion of power in the $4^{\circ}$ case is shown by the gray shaded region.}
\label{fig:cut_width}
\end{figure*}

We then ran the same ABC analysis for inferring the overall abundance of CDM subhalos using this GD-1 model and the posterior distribution and the predicted abundance are shown in Figure \ref{fig:prog-20_ABC_rate} and they are almost identical to those obtained with the fiducial GD-1 progenitor position. Therefore, the inferences in this paper do not change when we use the alternative progenitor position. The slightly higher abundance prediction in this model is expected as mentioned above.

\section{Adding perturbations at linear order}
\label{sec:add densities}

In this appendix, we demonstrate that the density power spectrum as a result of linearly adding the density perturbations due to the bar and spiral arms to the perturbed stream density due to impacts with subhalos, GMCs, and GCs agrees with that due to the combined effects of all the perturbers to good enough accuracy for the purposes of our analyses. We use the same implementation of the \textit{particle spray} method as described in \citet{Banik2019}, which is based on the technique from \citet{Fardal2014}. We first compute the unperturbed stream density $\Delta_{\rm{smooth}}$ by sampling the phase space coordinates and stripping time of $10^{5}$ points along the stream generated in the axisymmetric Milky Way potential, and binning them in $1^{\circ}$ bins along the length of the stream. To incorporate the effects of the bar and spiral arms, we integrate the same sampled points back to their respective time of stripping in the axisymmetric Milky Way potential and then integrate them forward in the bar + spiral Milky Way potential until today. We then bin them over the same bins and compute their density $\Delta_{\rm{bar+spiral}}$. We subtract $\Delta_{\rm{smooth}}$ from $\Delta_{\rm{bar+spiral}}$ to obtain the density perturbation $\delta_{\rm{bar+spiral}}$ due to the bar and spiral arms. Next, we incorporate the effects due to the subhalo + GMC + GC impacts by first computing the parameters of impact (perturber's mass, scale radius, flyby velocity, angle of impact, distance of closest approach) and the time of impact following the same procedure as described in sections \ref{sec:GMC_GC} and \ref{sec:subhalos} and computing the velocity offsets using the impulse approximation. We then integrate the same set of sampled points as for the bar+spiral perturbation back in time in the axisymmetric Milky Way potential until their respective time of stripping, followed by integrating them forward in the same potential until today while adding the velocity offsets to the points at their respective time of encounter with the perturbers. The points are then binned as before to compute the perturbed density $\Delta_{\rm{subhalo + GMC + GC}}$. Finally, to compute the perturbed density due to the cumulative effect of all the perturbers, $\Delta_{\rm{cumulative}}$, we follow the same procedure as for the subhalo + GMC + GC impacts, except that we integrate the points forward in the bar + spiral Milky Way potential. 

\begin{figure*}
\centering
\begin{minipage}{.5\textwidth}
  \centering
  \includegraphics[width=0.8\linewidth]{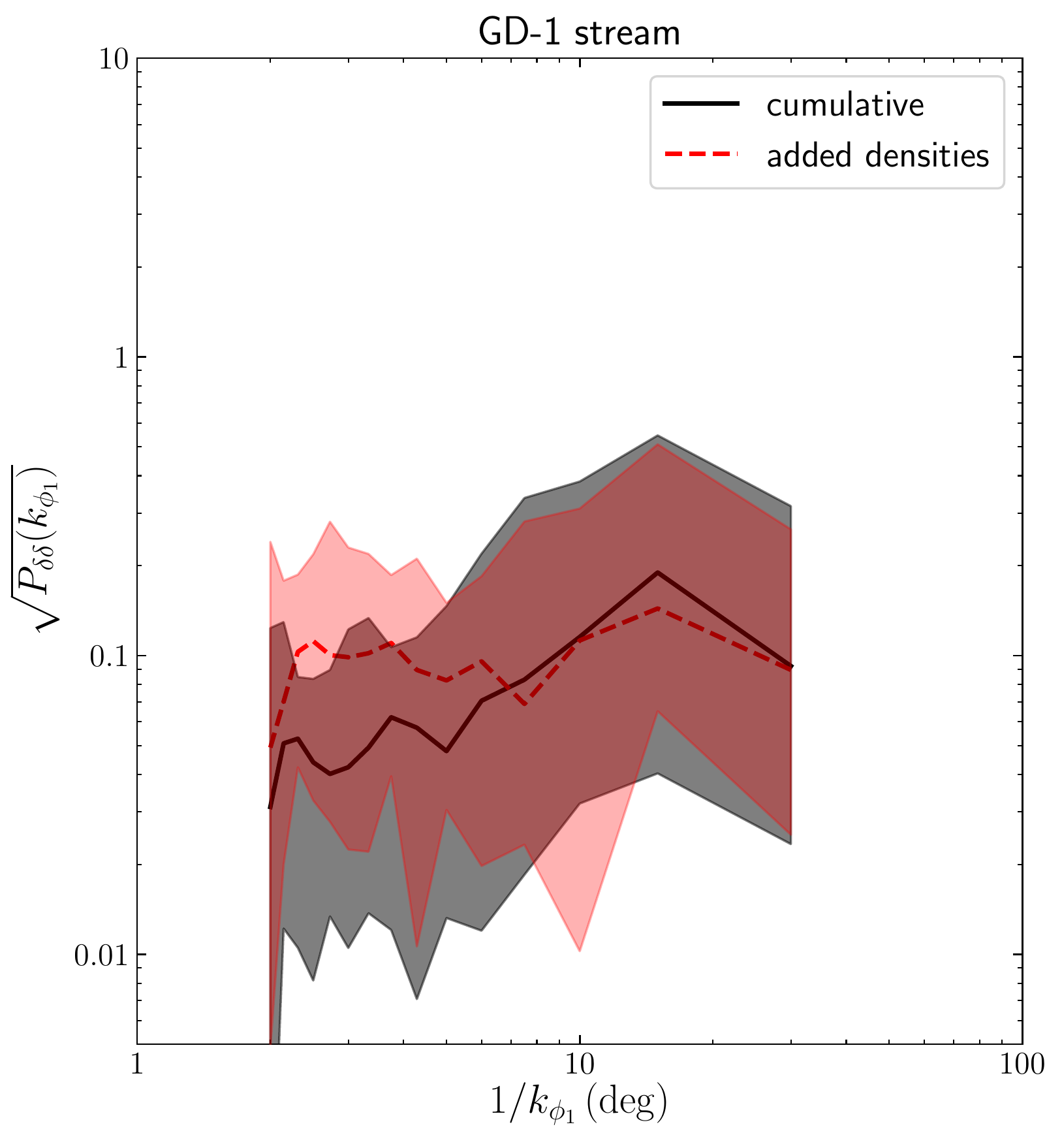}
  
  \end{minipage}%
\begin{minipage}{.5\textwidth}
  \centering
  \includegraphics[width=0.8\linewidth]{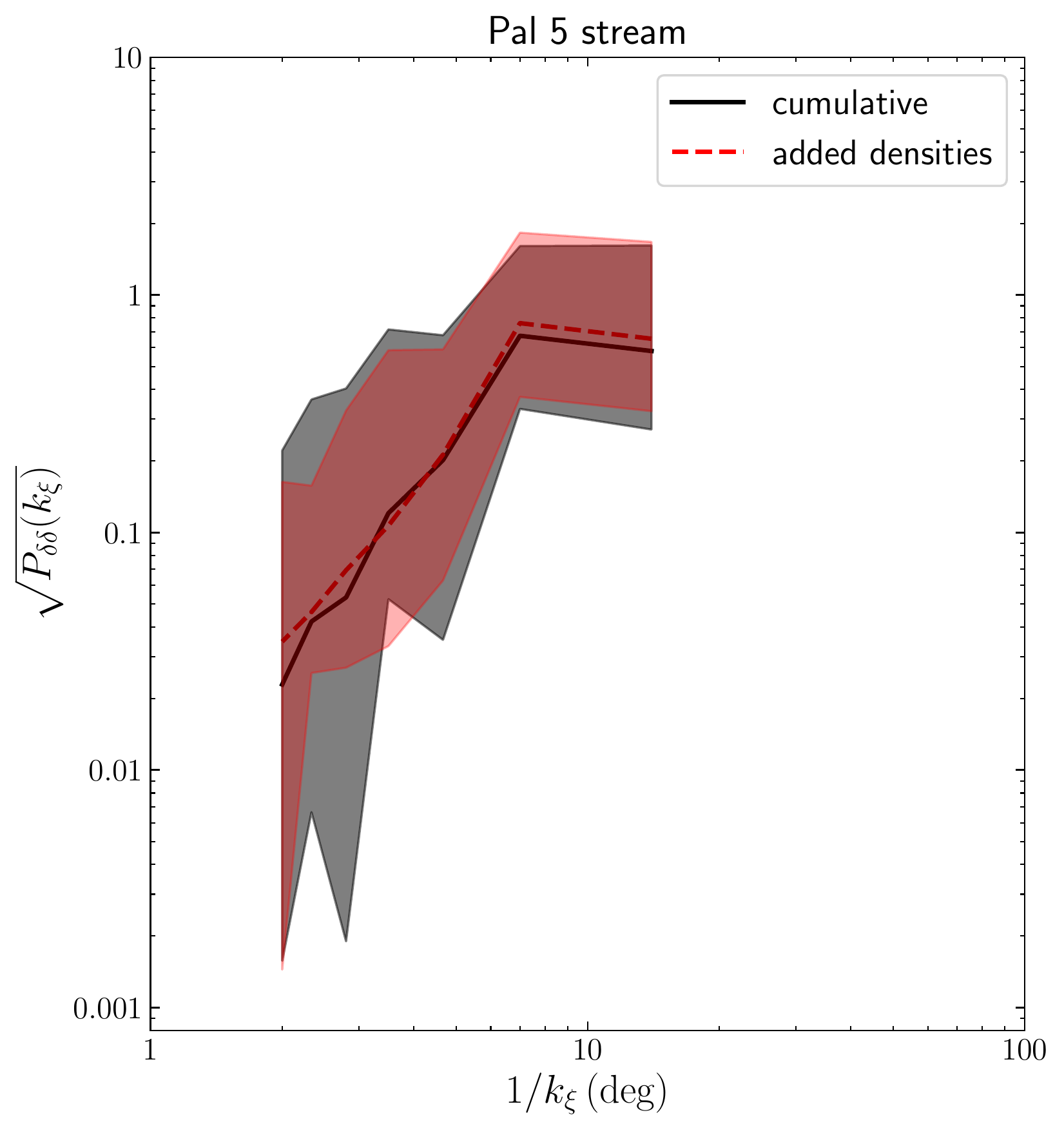}
\end{minipage}
\caption{Comparison of the density power spectrum as a result of adding the density perturbations due to the bar and spiral arms $\delta_{\rm{bar+spiral}}$ to the perturbed density due to the subhalo + GMC + GC impacts, $\Delta_{\rm{subhalo + GMC + GC}}$ for the GD-1 and Pal 5 streams. The red dashed line shows the median density power of the 20 different realizations with added density perturbations and the red shaded region shows the scatter. The black solid line shows the median density power of the 20 realizations with the cumulative effects of all the perturbers on the stream and the gray shaded region showing the corresponding scatter. The overlap of these two shaded regions appear to give rise to a third color which is only a plotting artifact. The left panel shows the case of a 7 Gyr old GD-1 stream while the right panel shows the case of the Pal 5 stream that is 5 Gyr old. Both median and scatter of the density power converges for both streams at scales where the signal dominates noise, demonstrating that we can compute perturbations due to the bar+spiral and the subhalos + GMCs + GCs separately and add them for the purpose of our ABC analyses in the main text.}
\label{fig:density_pk_convergence}
\end{figure*}

To linear order, we should have that $\Delta_{\rm{cumulative}} = \Delta_{\rm{bar+spiral}}+\Delta_{\rm{subhalo + GMC + GC}}$ and, most important for our analyses, that their power spectra agree. Having computed the density perturbations, we compute $\Delta_{\rm{subhalo + GMC + GC}} + \delta_{\rm{bar+spiral}}$ and compute its power spectrum and compare with that due to $\Delta_{\rm{cumulative}}$. We compute the scatter and median of the power spectra of 20 realizations of the trailing arm of the Pal 5 stream and 20 realizations of the trailing arm of a 7 Gyr GD-1 stream. We only test the oldest GD-1 stream model since it will have the highest value of density perturbation due to the bar and spiral arms $\delta_{\rm{bar+spiral}}$, and convergence of its power spectra would imply that the power spectra for the younger streams models will also converge. The results are shown in Figure \ref{fig:density_pk_convergence}.  The left panel shows the case for the GD-1 stream where we assumed the fiducial CDM subhalo abundance. The right panel shows the case for the Pal 5 stream for which we used 0.5 times the fiducial CDM subhalo abundance since for abundances higher than that majority of the stream realizations get completely disrupted. Also, as evident from Figure \ref{fig:ABC_rate}, relative abundances above 0.5 are strongly disfavored. The black solid lines represent the median power and the gray shaded regions show the scatter of the 20 different realizations in which the cumulative effects of all the perturbers namely, the bar, spiral arms, GMCs, GCs and subhalos are taken into account. Similarly, the red dashed lines represent the median and the red shaded region represent the scatter of the realizations in which the density perturbations due to the bar and spiral arms are added to the perturbed density due to impacts by the subhalos, GMCs and GCs. As evident, the median power converges at scales greater than $8^{\circ}$ in $\phi_{1}$ for GD-1 and above $4^{\circ}$ in $\xi$ for Pal 5, which are the relevant scales for the streams where the signal dominates noise. It is worth pointing out that although using the \textit{particle spray} technique we are able to simulate the combined effect of the bar, spiral arms, GMCs, GCs and subhalos, it is highly time consuming and hence we do not use it for our ABC calculations.

\end{document}